\documentclass[acmtog]{acmart} %

\acmSubmissionID{0194} %

\usepackage{import} %
\graphicspath{{figs/}}

\usepackage[ruled,noend]{algorithm2e} %

\SetAlFnt{\small}
\SetAlCapFnt{\small}
\SetAlCapNameFnt{\small}
\SetAlCapHSkip{0pt}
\IncMargin{-\parindent}
\usepackage{ifpdf}
\usepackage{graphicx}
\ifpdf
  \newcommand{\TODO}[3]{\textcolor{#2}{\textbf{[\textsc{#1:} \textit{#3}}]}}
\else
  \newcommand{\TODO}[3]{{\textbf{[\textsc{#1:} \textit{#3}}]}}  %
\fi
\usepackage{color}
\definecolor{cyan}{cmyk}{1,0,0,0}
\definecolor{darkred}{rgb}{0.7,0,0}
\definecolor{darkgreen}{rgb}{0,0.7,0}
\definecolor{darkblue}{rgb}{0,0,0.7}
\definecolor{orange}{rgb}{1,0.5,0}
\definecolor{magenta}{cmyk}{0,1,0,0}
\definecolor{darkyellow}{cmyk}{0,0,0.75,0}
\definecolor{gray}{rgb}{0.8,0.8,0.8}

\usepackage{amsmath} %

\usepackage[toc,page]{appendix} %
\usepackage{wrapfig} %
\usepackage{comment}
\usepackage{cuted} %
\usepackage{lipsum} %
\usepackage{mathtools} %
\usepackage{algorithmicx}
\usepackage{algpseudocode}

\usepackage{gensymb}
\usepackage{float}
\usepackage{soul}
\usepackage{array}
\usepackage{multirow}
\usepackage{hhline}
\usepackage{setspace}

\usepackage{hyperref} %

\algdef{SE}[DOWHILE]{Do}{doWhile}{\algorithmicdo}[1]{\algorithmicwhile\ #1}%

\makeatletter
\renewcommand{\ALG@beginalgorithmic}{\small}
\makeatother

\newcommand{\DELETE}[1]{} %
\newcommand{\IGNORE}[1]{}
\usepackage{datenumber}
\usepackage{calc}
\usepackage[mmddyyyy]{datetime}

\newcounter{datetoday}
\newcounter{diffyears}
\newcounter{diffmonths}
\newcounter{diffdays}

\newcommand{\difftoday}[3]{%
      \setmydatenumber{datetoday}{\the\year}{\the\month}{\the\day}%
      \setmydatenumber{diffdays}{#1}{#2}{#3}%
      \addtocounter{diffdays}{-\thedatetoday}%
      \ifnum\value{diffdays}>0
        \def\diffbefore{}%
        \def\diffafter{left}%
      \else
        \def\diffbefore{}%
        \def\diffafter{ago}%
        \setcounter{diffdays}{-\value{diffdays}}%
      \fi
      \setcounter{diffyears}{\value{diffdays}/365}%
      \setcounter{diffdays}{\value{diffdays}-365*\value{diffyears}}%
      \setcounter{diffmonths}{\value{diffdays}/30}%
      \setcounter{diffdays}{\value{diffdays}-30*\value{diffmonths}}%
      \diffbefore
      \ifnum\value{diffyears}=0
      \else
        \ifnum\value{diffyears}>1
            \thediffyears\space years,
        \else
            \thediffyears\space year,
        \fi
      \fi
      \ifnum\value{diffmonths}=0
      \else
        \ifnum\value{diffmonths}>1
            \thediffmonths\space months
        \else
            \thediffmonths\space month
        \fi
      \fi
      \ifnum\value{diffdays}=0
      \else
        \ifnum\value{diffdays}>1
            \thediffdays\space days
        \else
            \thediffdays\space day
        \fi
      \fi
      \diffafter
}

\def\thickhline{\noalign{\hrule height 1pt}}

\renewcommand{\TODO}[1]{}

\newcommand{\NEW}[1]{{\color{black}{#1}}} %
\newcommand{\NEWR}[1]{#1}%
\newcommand{\NEWRR}[1]{{\color{black}{#1}}}%

\def\norm#1{\left\|#1\right\|}
\def\d{\mathrm{d}}
\def\inv#1{\frac{1}{#1}}
\def\dfrac#1#2{\frac{\d #1}{\d #2}}
\def\pfrac#1#2{\frac{\partial #1}{\partial #2}}

\def\bfx{\mathbf x}
\def\bfy{\mathbf y}
\def\bfz{\mathbf z}
\def\bfn{\mathbf n}
\def\bfp{\mathbf p}

\def\calM{\mathcal{M}}
\def\calE{\mathcal{E}}
\def\calV{\mathcal{V}}
\def\calT{\mathcal{T}}
\def\calB{\mathcal{B}}

\def\R{\mathbb{R}}

\def\homega{\hat{\omega}}
\def\exbdM{ {\overline{\partial\mathcal{M}}} }

\def\sPath{\bar{\bfx}}
\def\sMatPath{\bar{\bfp}}

\def\sDelay{\mathbf{t}}
\def\sDelaysSpace{\mathcal{T}}
\def\ft{f_\mathrm{t}}
\def\ftau{f_\mathcal{T}}
\def\lrarrow{\leftrightarrow}
\def\tof{\mathrm{tof}}

\newcommand{\dn}[1]{\overset{\square}{#1}}
\newcommand{\fDiff}[2]{\frac{\mathrm{d}#1}{\mathrm{d}#2}}

\newcommand{\btheta}{\boldsymbol\uptheta}
\newcommand{\dF}{\dot{f}}

\newcommand{\dnF}{\overset{\square}{f}}

\newcommand{\dnFtau}{\overset{\square}{f}_\mathcal{T}}
\newcommand{\matFtau}{\hat{f}_\mathcal{T}}
\newcommand{\dmatFtau}{\left(\hat{f}_\mathcal{T}\right)^\cdot}
\newcommand{\dnmatFtau}{\left(\hat{f}_\mathcal{T}\right)^\square}

\newcommand{\sPathMap}{\hat{\bfx}}

\newcommand{\dnSe}{\overset{\square}{S}_e}

\newcommand{\dnT}{\overset{\square}{\mathfrak{T}}}

\newcommand{\dntof}{\overset{\square}{\tof}}

\newcommand{\boundary}[1]{\partial{\overline{#1}}}

\usepackage{booktabs} %

\usepackage[ruled]{algorithm2e} %
\usepackage[shortlabels]{enumitem}

\SetAlFnt{\small}
\SetAlCapFnt{\small}
\SetAlCapNameFnt{\small}
\SetAlCapHSkip{0pt}
\SetKwInput{KwData}{Input}
\SetKwInput{KwResult}{Output}

\setcopyright{rightsretained}
\acmJournal{TOG}
\acmYear{2021}\acmVolume{40}\acmNumber{6}\acmArticle{285}\acmMonth{12} \acmDOI{10.1145/3478513.3480498}

\begin{document}
\title{Differentiable Transient Rendering}

\author{Shinyoung Yi}
\author{Donggun Kim}
\author{Kiseok Choi}
\affiliation{%
  \institution{KAIST}
  \department{School of Computing}
  \city{Daejeon, South Korea}
  \postcode{34141}
}

\author{Adrian Jarabo}
\author{Diego Gutierrez}
\affiliation{%
  \institution{Universidad de Zaragoza, I3A}
  \department{}
  \city{Zaragoza}
  \postcode{50018}
}

\author{Min H. Kim}
\affiliation{%
  \institution{KAIST}
  \department{School of Computing}
  \city{Daejeon, South Korea}
  \postcode{34141}
}
\email{corresponding_author: minhkim@kaist.ac.kr}

\renewcommand{\shortauthors}{Yi, Kim, Choi, Jarabo, Gutierrez, and Kim}

\begin{abstract}
Recent differentiable rendering techniques have become key tools to tackle many inverse problems in graphics and vision. Existing models, however, assume steady-state light transport, i.e., infinite speed of light. While this is a safe assumption for many applications, recent advances in ultrafast imaging leverage the wealth of information that can be extracted from the exact time of flight of light. In this context, physically-based transient rendering allows to efficiently simulate and analyze light transport considering that the speed of light is indeed finite. In this paper, we introduce a novel differentiable transient rendering framework, to help bring the potential of differentiable approaches into the transient regime. To differentiate the transient path integral we need to take into account that scattering events at path vertices are no longer independent; instead, tracking the time of flight of light requires treating such scattering events at path vertices jointly as a multidimensional, evolving manifold. We thus turn to the generalized transport theorem, and introduce a novel \textit{correlated importance} term, which links the time-integrated contribution of a path to its light throughput, and allows us to handle discontinuities in the light and sensor functions. Last, we present results in several challenging scenarios where the time of flight of light plays an important role such as optimizing indices of refraction, non-line-of-sight tracking with nonplanar relay walls, and non-line-of-sight tracking around two corners. 
\end{abstract}



\begin{teaserfigure}
	\centering
	\vspace{0mm}%
	\includegraphics[width=\linewidth]{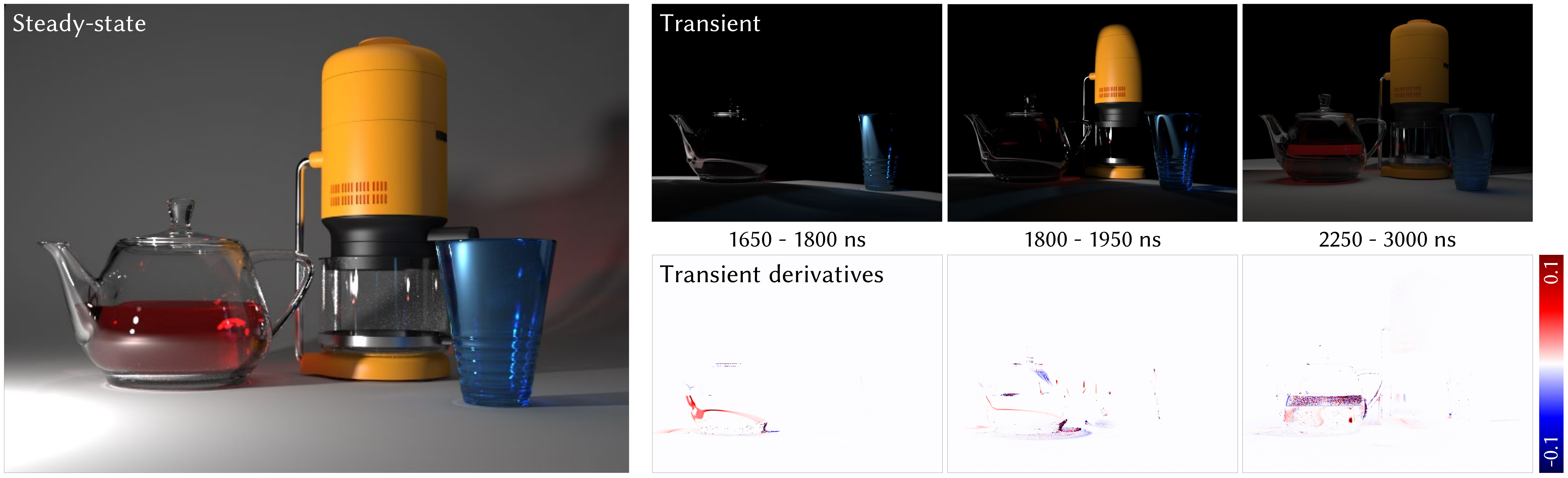}%
	\vspace{-2mm}%
	\caption[]{\label{fig:teaser}%
Our general-purpose differentiable transient rendering framework allows to compute derivates of complex, multi-bounce transient sequences with respect to scene parameters, even in the presence of discontinuous light and sensor functions. The figure shows steady-state and transient renders of a table-top scene with light coming from the left, then being bounced back by two off-camera diffusers. \NEW{The bottom row shows the transient light transport scene derivatives with respect to  the index of refraction of the red tea}. %
Please refer to Figure~\ref{fig:transparent-rendering} for additional results optimizing the index of refraction of the tea in the teapot, which in turn changes the speed of light. }
	\vspace{2mm}%
\end{teaserfigure}

\maketitle

\section{Introduction}
\label{sec:intro}

\NEW{Physically-based differentiable rendering deals with the computation of the derivatives of radiometric measurements, according to changes in scene parameters (see recent references~\cite{Zhao2020,kato2020differentiable}  for a wide overview of the field). It has recently become a key tool not only for inverse rendering or scene reconstruction problems that require gradient-based optimization, but also to enable the integration of physics-based simulations in machine learning pipelines, computing the loss function in rendering space}. While many different approaches have been presented in recent years (e.g.,~\cite{li2018differentiable, zhang2019differential, zhang2020path}), all of them assume \textit{steady-state} configurations, where time delays due to the propagation of light are ignored. 

\NEW{Transient imaging, on the other hand, has already enabled many non-trivial applications in computer vision and scene recognition \cite{jarabo2017recent}, such as visualizing light in motion \cite{velten2013femto,heide2013low}, looking around corners \cite{otoole2018confocal,liu2019phasor}, looking through turbid media \cite{heide2014imaging,wu2018adaptive}, or the decomposition of global illumination components \cite{wu2014decomposing, marco2021virtual}, to name a few. Transient rendering techniques, which lift the assumption that light speed is infinite and take explicitly into account light's time of flight,  have thus become an increasingly relevant tool to accurately simulate scenarios and hypotheses without requiring cumbersome data capture sessions, as well as to help provide rigorous mathematical analyses~\cite{jarabo2014framework,pediredla2019ellipsoidal}. }

Computing the time-resolved radiance derivatives as a function of the scene is currently limited to expensive and fragile finite differences~\cite{gkioulekas2016evaluation,iseringhausen2020non}, which hinders the applicability of current transient rendering in inverse problems or machine learning approaches. 
To overcome this, we introduce \textit{path-space differentiable transient rendering}. Our goal is to  differentiate the transient path integral ~\cite{jarabo2014framework}  to help unlock the potential of differential approaches in transient scenarios.
We make the observation that, when tracking the time of flight of light, scattering events at path vertices are no longer independent, as assumed in recent work~\cite{zhang2020path}.
As a consequence, we are no longer dealing with two-dimensional evolving manifolds, which in turn means that we cannot apply the Reynolds transport relation recursively between path vertices~\cite{zhang2020path}. 

We thus rely on the \textit{generalized transport theorem}~\cite{seguin2014roughening}, applied on the \textit{higher-dimensional manifold} that results from all path vertices contributing to the temporal domain, and introduce a novel \textit{correlated importance} term, which links the time-integrated contribution of a path to its light throughput and allows us to handle discontinuous light and sensor functions.  
From our theoretical formulation, we then demonstrate that, despite the higher dimensionality of transient light transport, our practical formulation of the boundary integral converges to Zhang's original formulation of differentiable steady-state path tracing~\shortcite{zhang2020path}. In that sense, our work can also be seen as a generalization of the existing path-space formulation, lifting the steady-state assumption of independent scattering events while being able to handle discontinuities in the light and sensor functions. In addition, we show how to incorporate our work in a Monte Carlo framework, defining estimators for both the interior and the boundary integrals. 

We validate our results against the baseline transient path integral technique and using finite differences. In addition, we demonstrate how our formulation enables several novel applications in challenging scenarios, where the time of flight of light plays an important role and traditional non-line-of-sight (NLOS) imaging has been inapplicable. In particular, we show optimization results for indices of refraction; NLOS tracking with a nonplanar, wavy relay wall; and tracking the motion of a hidden object around two corners. \NEW{These last two examples are the first demonstrations of NLOS imaging under such challenging scenarios, and demonstrate the potential of differential transient rendering for solving difficult inverse problems. We hope that our code and datasets\footnote{\url{http:/vclab.kaist.ac.kr/siggraphasia2021/}} will help future developments on the field.}

\section{Related Work}
\label{sec:relatedwork}

For a wide overview of the applications of transient imaging, we refer the reader to existing surveys on the topic~\cite{jarabo2017recent,satat2016advances}. In the following, we discuss related works that focus on transient and differentiable rendering.

\paragraph{Transient rendering}
The term "transient rendering" was coined by Smith et al.~\shortcite{smith2008transient}, who proposed to extend the rendering equation by including the temporal delay of light propagation. Based on this framework, several works proposed transient variations of Monte Carlo rendering~\cite{jarabo2012femto,krivanek2013recent, pitts2014time,ament2014refractive,otoole2014temporal,adam2015bayesian}. 
Jarabo et al.~\shortcite{jarabo2014framework} presented a practical framework by introducing the \textit{transient path integral}, including an efficient reconstruction technique, as well as a novel sampling strategy by adding indirect shadow vertices, to control path lengths in participating media. This framework was later extended to include vector-based effects such as polarization or fluorescence~\cite{jarabo2018bidirectional}.
Pediredla et al.~\shortcite{pediredla2019ellipsoidal} introduced ellipsoidal connections for time-gated rendering, also using additional vertices to control path length, but for surfaces instead of media.
Other authors have recently focused on increasing efficiency when rendering transient light transport, usually based on simplified models~\cite{iseringhausen2020non,tsai2019beyond,pan2019transient,chen2020learned}, or by means of filtering~\cite{marco2019progressive}. 
All these works render time-resolved zeroth-order radiance; in contrast, we develop a differentiable transient path integral, and focus on the derivatives of transient light transport.

\paragraph{Analysis-by-synthesis in transient imaging}
Several works have leveraged transient rendering for inverse problems, using it as the generative model for gradient descent-based optimization. Iseringhausen and Hullin~\shortcite{iseringhausen2020non} used an efficient three-bounce renderer for non-line-of-sight (NLOS) reconstruction based on finite differences, while Tsai et al.~\shortcite{tsai2019beyond} explicitly derived the light transport gradients of hidden surface properties. None of these works account for visibility changes, and are limited to the particular problem of three-bounce NLOS reconstruction. Gkioulekas et al.~\shortcite{gkioulekas2016evaluation} analyzed the use of transient rendering in the context of stochastic gradient descent for recovering heterogeneous media.
In contrast, our framework is general and does not rely on explicitly computed gradients, while taking into account complex derivatives including singularities produced by, e.g., visibility changes.

\paragraph{Differentiable rendering} 
While several special-purpose differentiable rendering systems have been proposed in the past (e.g.,~\cite{gkioulekas2016evaluation, kato2018neural, chen2019learning, liu2019soft}) in this work we focus on general-purpose differentiable rendering. 
It was first proposed in OpenDR~\cite{Loper2014opendr}, targeting scalability and efficiency; however, it was based on a simple physical model, which reduced its applicability to direct illumination only. Laine et al.~\shortcite{laine2020modular} included additional features such as antialiasing or texture filtering, but still limited only to direct illumination. 

Li et al.~\shortcite{li2018differentiable} proposed a general-purpose, differentiable, and physically-based rendering model, computing the differential form of the rendering equation with Monte Carlo ray tracing, while taking into account global illumination and geometric discontinuities. The authors proposed searching for the discontinuities on the integral using edge sampling. The discontinuity handling was later improved by the works of Loubet et al.~\shortcite{loubet2019reparameterizing}, who reparameterized the integral on top of the differentiable renderer Mitsuba 2~\cite{nimierdavid2019mitsuba} to ease computations at the price of introducing bias, and Bangaru et al.~\shortcite{bangaru2020unbiased}, who combined edge and area sampling for increasing robustness and efficiency. Zhang et al.~\shortcite{zhang2019differential} extended the differential rendering equation to the volumetric rendering equation, modeling light transport in participating media. These approaches work on the local domain of the rendering equation, which makes it difficult to extend them to  bidirectional rendering algorithms. 
Zhang et al.~\shortcite{zhang2020path} introduced a differential path-space light transport integral suitable for bidirectional methods. Our work builds on top of such integral, and generalizes it to time-resolved light transport. Finally, orthogonal to these works, Nimier-David et al.~\shortcite{nimierdavid2020radiative} proposed an adjoint differential light transport formulation that propagates the derivatives from the camera to the light sources, dramatically improving speed and memory requirements.

\begin{table}[]\small
\caption{\label{tb:symbols} 
	Main symbols used in the paper.}
\vspace{-3mm}
\begin{tabular}{cl}
\hline
Symbol   & Description \\ \hline
$\sPath=\bfx_0...\bfx_k$   & Light path of $k+1$ vertices\\
$\sDelay=t_0...t_k$   & Time delays on $k+1$ vertices \\

$\Omega$ & Space of all light paths\\
$\Omega_k$ & Space of light paths of $k+1$ vertices\\
$\mathcal{T}$ & Space of temporal delays \\

\hline
$c$ & Speed of light in vacuum \\
$\eta_i$ & Refractive index of the medium between $\bfx_i$ and $\bfx_{i-1}$\\
$\mathrm{tof}\left(\sPath\right)$ & Total delay of path $\sPath$ \\
$L_e\left(\bfx_0\to\bfx_1,t\right)$ & Light source emission function \\&{with direction $\bfx_0\to\bfx_1$, time delay $t$} \\
$W_e\left(\bfx_{k-1}\to\bfx_k,t\right)$ & Sensor sensitivity function\\&{with direction $\bfx_{k-1}\to\bfx_k$, time delay $t$} \\
$S_e\left(\sPath\right)$ & Correlated importance as a function of light path {$\sPath$} \\
$\rho\left(\bfx_i,t_i\right)$ & Scattering function at vertex $\bfx_i$\\
$G\left(\bfx_i\leftrightarrow \bfx_{i+1}\right)$ & Geometric function \\
$V\left(\bfx_i\leftrightarrow \bfx_{i+1}\right)$ & Visibility function \\
$\mathfrak{T}\left(\sPath,\sDelay\right)$& Light throughput, the product of $\rho$, $G$, and $V$\\

\hline
$\btheta=\theta_1...\theta_d$ & Set of $d$ parameters describing a scene\\
$\calM\left(\btheta\right)$ & Manifold that evolves with $\btheta$ \\
$\partial\calM\left(\btheta\right)$ & Boundary of $\calM${$\left(\btheta\right)$}\\
$v\left(\bfx\right)$ & Local velocity {at $\bfx$} \\
$\mathcal{V}_\calM\left(\bfx\right)$ & {Scalar} normal velocity of $\calM\left(\btheta\right)$ at $\bfx$ \\
{$\mathcal{V}_{\boundary{\calM}}\left(\bfx\right)$} & {Scalar} normal velocity of {$\boundary{\calM\left(\btheta\right)}$} at $\bfx$ \\
$\kappa\left(\bfx\right)$ & Total curvature {at $\bfx$} \\

\hline
$I(\btheta)$ & Rendered image as a function of $\btheta$\\
$f(\sPath)$ & Contribution to $I$ of path $\sPath$ \\
$\ft(\sPath,\sDelay)$ & Time-resolved contribution to $I$ of path $\sPath$ \\&with time delay $\sDelay$\\
$\ftau(\sPath)$ & Time-integrated contribution to $I$ of path $\sPath$\\
$\dot f_{\calT}\left(\sPath,\btheta\right)$ & {Derivative} of $f_{\calT}$ for path $\sPath$ w.r.t. $\btheta$ \\
$\overset{\square}{f_{\calT}}\left(\sPath,\btheta\right)$ & Normal {derivative} of $f_{\calT}$ for $\sPath$ w.r.t. $\btheta$\\
$\Delta {f_\calT}(\sPath)$ & Boundary contribution of function $f_{\calT}$ for $\sPath$ on $\boundary{\Omega}$\\
$\Delta\Omega\left[f_{\calT}\right]\left(\btheta\right)$ & Set of discontinuities of function $f_{\calT}$ on $\Omega$ w.r.t. $\btheta$\\
$\boundary{\Omega}\left[f_\calT\right]\left(\btheta\right)$ & Extended boundary of $\Omega$ w.r.t. $f_{\calT}$ and $\btheta$\\
{$\bfn(\bfx_i)$} & {Unit normal at $\bfx_i$}\\
{$\delta\left(t_i\right)$} & {Dirac delta at $t_i$}\\
{$\norm{\bfx_i-\bfx_{i-1}}^\square$} & {Normal derivative of $\norm{\bfx_i-\bfx_{i-1}}$ w.r.t. $\btheta$}\\
\hline
\end{tabular}

\vspace{-6mm} 
\end{table}

\section{Background}
\label{sec:background}

We introduce here the main aspects of both the transient path integral and path-space differentiable rendering. Table \ref{tb:symbols} summarizes the main symbols used throughout the paper. 

\subsection{Transient Path Integral}
\label{sec:transientrendering}

The path integral~\cite{veach1997robust} models the intensity $I$ recorded by a virtual sensor as the integral of all light paths $\Omega$ contributing to a pixel, assuming that both the emission and sensor response are orders of magnitude larger than the propagation time of light. 
Incorporating the temporal domain we obtain the \textit{transient path integral} ~\cite{jarabo2014framework} as
\begin{equation}
\label{eq:pathintegral}
I = \int_\Omega\int_\sDelaysSpace \ft(\sPath,\sDelay) \d\mu(\sDelay) \d\mu(\sPath) ,
\end{equation}
where $\sDelaysSpace$ represents the space of temporal delays, $\sDelay=t_0...t_k$ is the sequence of time delays on each vertex, $\d\mu(\sDelay)$ denotes temporal integration at each vertex,  $\d\mu(\sPath)$ is the differential measure, and $\sPath=\bfx_0...\bfx_k$ is a path of $k+1$ vertices. Vertices $\bfx_0$ and $\bfx_k$ lie on the light source and the sensor, respectively.
For convenience, we define the path space as $\Omega = \cup_{k=1}^\infty \Omega_k$, with $\Omega_k$ being the space of all paths with $k$ vertices. The path contribution $\ft(\sPath,\sDelay)$ is given by
\begin{equation}
\label{eq:tran_path_integ}
\ft(\sPath,\sDelay) = L_e(\bfx_0 \to \bfx_1, t_0)\, \mathfrak{T}(\sPath,\sDelay)\, W_e(\bfx_{k-1}\to\bfx_k, t_k+\tof(\sPath)),
\end{equation}
where $\tof(\sPath)=c^{-1}\sum_{i=0}^{k-1}||\bfx_i-\bfx_{i-1}||\eta_i + \sum_{i=0}^{k-1}t_i$ is the total delay of path $\sPath$ (i.e., its time of flight), with $c$ the speed of light and $\eta_i$ the index of refraction of the medium between vertices $\bfx_i$ and $\bfx_{i+1}$. 
The temporal delay on the emission $L_e(\bfx_0 \to \bfx_1, t_0)$ is continuous in real-world applications, ranging from a few femto-seconds in ultrashort lasers to continuous light modulation in time-of-flight cameras. The sensor sensitivity $W_e(\bfx_{k-1}\to\bfx_k, t_k+\tof(\sPath))$ models the temporal response of each specific timestamp, in addition to the spatio-angular response of the pixel. 
Last, the light throughput $\mathfrak{T}(\sPath,\sDelay)$ is obtained from the product of the scattering function at inner vertices $\rho$, the geometry function $G$, and the visibility term $V$ for each path segment as
\begin{equation}
\label{eq:tran_path_throughput}
	\mathfrak{T}(\sPath,\sDelay) = \left[
	\prod_{i=1}^{k-1}\rho\left( \bfx_i,t_i\right)
	\right]\left[
	\prod_{i=0}^{k-1}G\left(\bfx_i \lrarrow \bfx_{i+1}\right) V\left(\bfx_i \lrarrow \bfx_{i+1}\right)
	\right], 
\end{equation}
where the time dependency is explicitly incorporated in the scattering function as $\rho\left( \bfx_i,t_i\right)$.

\subsection{Path-space Differentiable Rendering}
\label{sec:diffrendering}

Differentiable rendering deals with the computation of the derivatives of the image $I(\btheta)$ with respect to a set of parameters $\btheta$ describing the scene. Unfortunately, the path space $\Omega(\btheta)$ presents a large number of discontinuities due to occlusions and surface discontinuities, resulting into potential interactions between the different parameters; a na\"{i}ve derivation of $\fDiff{I(\btheta)}{\btheta}$ cannot handle such \NEW{\emph{evolving} discontinuities, and therefore require sophisticated methods to handle them~\cite{li2018differentiable,loubet2019reparameterizing,bangaru2020unbiased}.}

Zhang et al.~\shortcite{zhang2020path} observed that, 
given a fixed set of parameters~$\btheta$, the vertices $\bfx_{0...k}$ of a path $\sPath\in\Omega(\btheta)$ lie on a surface manifold $\calM(\btheta)\subset\R^2$ that \emph{evolves} with $\btheta$. In particular, it follows a vector field defined by the local velocities $v(\bfx_{0...k},\btheta)=\fDiff{}{\btheta}\bfx_{0...k}(\btheta)$, with $\bfx_{0...k}(\btheta)\in\calM(\btheta)$ a path vertex for a parameter set $\btheta$ (in the following, we omit the dependence on $\btheta$ on paths and vertices for clarity). 

The manifold $\calM(\btheta)$ can be assumed C$^0$-continuous except in the set $\boundary{\calM(\btheta)}[f]\subset\calM(\btheta)$ that includes the boundaries of $\calM(\btheta)$ and the discontinuity curves of $f(\sPath)$, i.e., the jump discontinuity points in $f(\sPath)$, $\partial\calM$, and $\Delta\calM[f]$. For clarity, in the following we omit the dependence on $f$ of the extended boundary. 
We denote $\boundary{\Omega(\btheta)}\subset\Omega(\btheta)$ the subspace of paths with at least one vertex $\bfx_{0..k}\in\boundary{\calM(\btheta)}$, i.e., lying on the boundary manifold. Tracking this particular subset of the manifold allows to explicitly account for the derivatives due to discontinuities in the path space. 

\begin{figure}[tp]
	\centering
	\includegraphics[width=\linewidth]{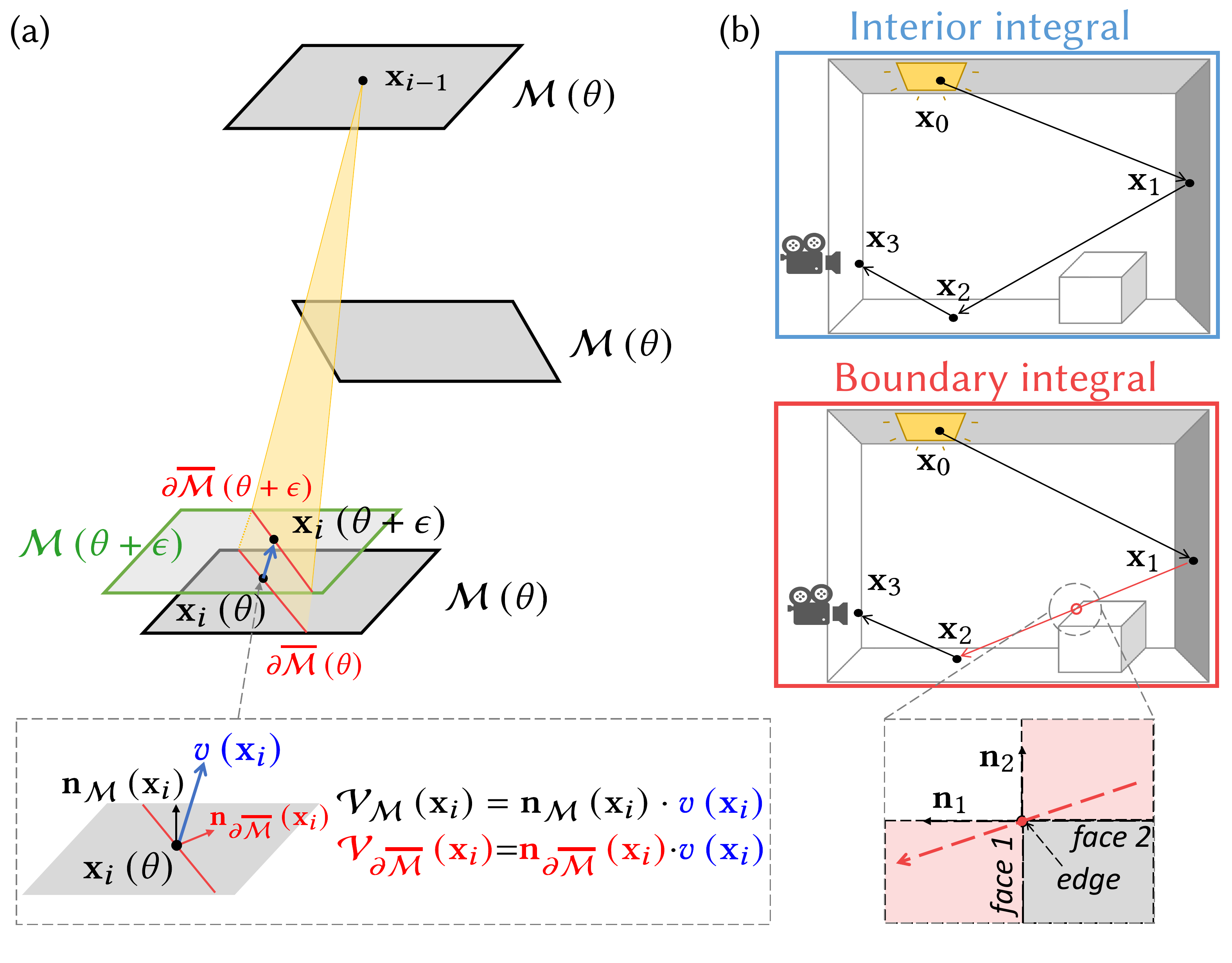}%
	\vspace{-2.5mm}%
	\caption[]{\label{fig:interior_boundary_sampling}
(a) A surface manifold $\calM(\theta)$ and a scalar field with a jump discontinuity evolve with respect to a scene parameter $\theta$. The red line represents the discontinuity of the scalar field (e.g., the visibility function $V(\bfx_i,\bfx_{i-1})$ for a fixed $\bfx_{i-1}$). Such discontinuity, belonging to  $\partial\overline{\calM}$, also evolves with~$\theta$. The blue arrow indicates the local velocity of a point $\bfx_i \in \partial\overline{\calM}$, denoted by $v\left(\bfx_i\right)$. 
				(b) Illustration of the sampling process of the interior integral (top) and the boundary integral (bottom) in path-space differentiable rendering~\cite{zhang2020path}. In the latter, light paths should contain a boundary segment (in red), which grazes a sharp edge in the scene geometry. 
	}
	\vspace{-3mm}
\end{figure}

By following the Reynolds transport theorem~\cite{cermelli2005transport} from fluid mechanics, the derivatives of the path integral can be computed by defining the integral on the evolving manifold $\calM(\btheta)$ as the sum of the so-called \emph{interior} and \emph{boundary} integrals, as 
\begin{align}
	\label{eq:ss_diff_path_integ}
\fDiff{}{\btheta}I(\btheta) & = \int_{\Omega}\left[\dnF(\sPath)-f(\sPath)\sum_{i=0}^k \kappa(\bfx_i) \calV_{\calM}(\bfx_i) \right] \d\mu(\sPath)  \\ \nonumber
& + \int_{\boundary{\Omega}} \Delta{f_i}(\sPath) \calV_{\boundary{\calM}_i}(\bfx_i) \d\mu_{\boundary{\Omega}}(\sPath),
\end{align}
where $\dnF(\sPath)$ is the {normal} derivative of $f(\sPath)$ as a function of~$\btheta$, 
$\Delta {f_i}(\sPath)$ is the boundary contribution function for a vertex $\bfx_i\in\boundary{\calM}$ on the boundary, 
$\kappa(\bfx_i)$ is the total curvature at $\bfx_i$, and $\calV_{\calM}(\bfx_i)$ (and $\calV_{\boundary{\calM}}(\bfx_i)$) are the scalar normal velocities 
defined as $\calV_{\calM}(\bfx_i)=\bfn(\bfx_i)\cdot v(\bfx_i)$ with $\bfn(\bfx_i)$ the unit normal at $\bfx_i$. 
The {normal} derivative of $f(\sPath)$ is defined as $\dnF(\sPath)=\dF(\sPath) - (v(\sPath)-\calV_{\calM} \bfn))\, \mathrm{grad}_\calM\left(f(\sPath)\right)$, with $\mathrm{grad}_\calM\left(f(\sPath)\right)$ the manifold gradient of $f(\sPath)$ (see Figure~\ref{fig:interior_boundary_sampling}.a). Note that for zero tangential velocity of the evolving manifold with respect to the local parameterization
we get $\dnF(\sPath)=\dF(\sPath)$.
Figure~\ref{fig:interior_boundary_sampling}.b depicts the derivatives of the path integral as interior and boundary integrals.

\def\norm#1{\left\|#1\right\|}
\def\d{\mathrm{d}}
\def\inv#1{\frac{1}{#1}}
\def\dfrac#1#2{\frac{\d #1}{\d #2}}
\def\pfrac#1#2{\frac{\partial #1}{\partial #2}}

\def\bfx{\mathbf x}
\def\bfy{\mathbf y}
\def\bfz{\mathbf z}

\def\calM{\mathcal{M}}
\def\calE{\mathcal{E}}

\def\R{\mathbb{R}}

\def\bfx{\mathbf x}
\def\bfy{\mathbf y}
\def\bfz{\mathbf z}

\def\homega{\hat{\omega}}
\def\exbdM{ {\overline{\partial\mathcal{M}}} }
\def\bdps{ {\overline{\partial\Omega}} }

\section{Differentiable Transient Rendering}
\label{sec:ourmethod}
In this section we obtain the differentiable version of the transient path integral in Equation~\eqref{eq:pathintegral}.
Following the path-space model proposed by Zhang et al.~\shortcite{zhang2020path},	 we compute two different terms for the path integral, the interior and boundary terms. 

In the original \textit{steady-state} formulation of Zhang and colleagues, all scattering events in path $\sPath$ are considered \textit{independent}, which allows the authors to apply the two-dimensional Reynolds transport theorem recursively. 
However, in \textit{transient state} this is no longer the case; the total time of flight of the path $\tof(\sPath)$ needs to be taken into account, and as a consequence scattering events are no longer independent. To deal with this and derive our differential transient path integral, we rely on the \textit{generalized transport theorem}~\cite{seguin2014roughening} applied on a higher-dimensional manifold.

\paragraph{Differential transient path integral}

\begin{figure}[t]
	\centering
	\includegraphics[width=\linewidth]{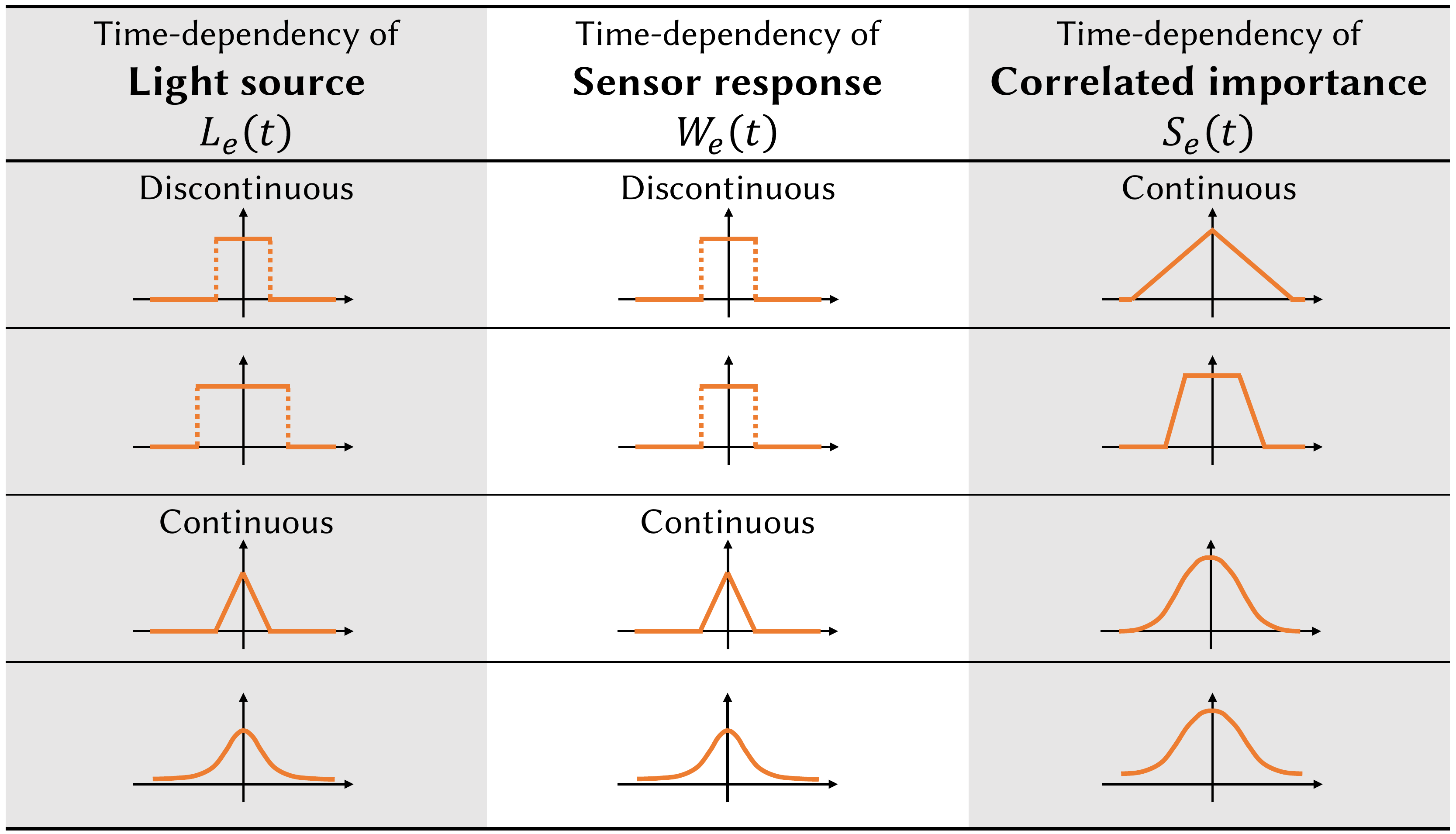}%
	\vspace{-2.5mm}%
	\caption[]{\label{fig:case_of_temporal_sensitivity}
		Example shapes of time-dependent light source $L_e$, sensor response $W_e$, and the resulting correlated importance $S_{e}$ functions. \NEW{These shapes include combinations of short square wave laser pulses and time-gated sensors, as well as Gaussian laser pulses and time-gated sensors, used in real applications.}}
	\vspace{-3mm}
\end{figure}

We first  define an intermediate path contribution term $\ftau$ that only depends on spatial variables as
\begin{equation}
	\label{eq:trans_path_contir_spatial}
	\ftau(\sPath)= \int_\sDelaysSpace \ft(\sPath,\sDelay)\d\mu(\sDelay), 
\end{equation}
which leads to the following expression for the transient path integral defined in Equation~\eqref{eq:pathintegral}:
\begin{equation}
\label{eq:altpathintegral}
I = \int_\Omega\ftau(\sPath)\d\mu(\sPath).
\end{equation}
\NEW{Similar to previous works~\cite{jarabo2014framework,tsai2019beyond,chen2020learned}, we assume that scattering delays in materials (due to e.g., multiple scattering or electromagnetic phase shift) are negligible compared with the temporal resolution of sensors and the propagation delays.
This allows us to approximate $\rho(\bfx_i,t_i)\approx\rho(\bfx_i)\delta(t_i)$, therefore removing the temporal dependence on the light throughput $\mathfrak{T}(\sPath)$ defined in Equation \eqref{eq:tran_path_throughput}.} 
Moreover, the integral domain of Equation~\eqref{eq:trans_path_contir_spatial} no longer depends on scattering delays, and thus it reduces to a one-dimensional time domain.
Therefore, when the temporal response of the source $L_e$ and sensor $W_e$ are independent,
the path contribution $\ftau$ becomes
\begin{align}
	\label{eq:trans_path_contir_nodelay}
	\ftau(\sPath) & = \mathfrak{T}(\sPath)\, \int_{-\infty}^{\infty}
L_e(\bfx_0 \to \bfx_1, t) W_e(\bfx_{k-1}\to\bfx_k, t+\tof(\sPath)) \d t \,  \nonumber \\
& =  \mathfrak{T}(\sPath)\,S_e(\sPath),
\end{align}
where we introduce a novel \textit{correlated importance} function $S_e(\sPath)$. 
Note that $S_e(\sPath)$ has an implicit dependence on time through $\tof(\sPath)$; this dependence will be relevant when we obtain its derivatives later in this section, since it allows us to handle discontinuous light and sensor functions (see Figure~\ref{fig:case_of_temporal_sensitivity}).

To compute the derivative of image $I$ in Equation~\eqref{eq:altpathintegral} we need to differentiate a high-dimensional integral, 
since our evolving manifolds are $2(k+1)$-dimensional in path space $\Omega$; more generally, $\Omega$ is an $m$-dimensional subset of $\R^n$, with $m$ and $n$ integers such that $n>m$.
However, due to the influence of $\tof(\sPath)$  in path contribution $f_{\calT}(\sPath)$, we cannot separate it into two-dimensional integrals on $\calM$ as in previous work~\cite{zhang2020path}, and as a consequence we cannot recursively apply Reynolds transport theorem for two-dimensional manifolds.

\NEW{Instead, we turn to the \emph{generalized transport theorem}~\cite{seguin2014roughening} to compute the derivative of $I$. As opposed to Reynolds transport theorem, the generalized transport theorem considers a higher-dimensional manifold $\Omega$ defined as the product of multiple two-dimensional manifolds. Intuitively, each of these low-dimensional manifolds define a light interaction at each path vertex. 
Applying the generalized transport theorem allows to directly compute the derivative of the full path, leading to our \textit{differential transient path integral} (see the supplemental document for the complete rigorous derivation): }
\begin{align}
	\label{eq:diff_trans_path_int}
	\fDiff{}{\btheta} \int_{\Omega} \ftau(\sPath) \d \mu(\sPath)
	& = \int_{\Omega}\left[\dnFtau(\sPath)-\ftau(\sPath)\sum_{i=0}^k \kappa(\bfx_i) \calV_{\calM}(\bfx_i) \right]\, \d \mu(\sPath) \nonumber\\ 
	& + \int_{\boundary{\Omega}} \Delta\ftau(\sPath)  \calV_{\boundary{\calM}}(\sPath) \,\d \mu_{\boundary{\Omega}}(\sPath),
\end{align}
where the first term represents the interior integral, and the second describes the boundary integral. Note that the boundary integral is now defined over the extended boundary 
$\boundary{\Omega}[\ftau]$; for clarity on the notation, we omit this dependence. %
Equation~\eqref{eq:diff_trans_path_int} has a similar structure to Zhang et al.'s steady-state differential path integral~\eqref{eq:ss_diff_path_integ}, but the term $\ftau(\sPath)$ now takes explicitly into account the temporal delays due to light's time of flight. Our transient formulation can thus be seen as a generalization of Zhang's formulation, but not limited to 2D manifolds. 
In the following, we derive individually the interior and boundary integrals.

\paragraph{Interior term}

From Equations \eqref{eq:trans_path_contir_nodelay} and~\eqref{eq:diff_trans_path_int}, the normal derivative of path contribution $\ftau(\sPath)$ with respect to $\btheta$ is
\begin{equation}
	\label{eq:diff_fcalt_twoterms}
	\dnFtau(\sPath) = \dnT(\sPath) S_e(\sPath) + \mathfrak{T}(\sPath) \dnSe(\sPath) .
\end{equation}
\NEW{
The normal derivative of the light throughput $\dnT\left(\sPath\right)$ is obtained by the product rule in a similar fashion as Zhang et al.:
	\begin{equation}\small
		\label{eq:deriv_throughput}
		\dnT(\sPath) = \sum_{j=1}^{k-1}\left[
			\dn{\rho}_j	\prod_{\substack{i=1\\i \ne j}}^{k-1}\rho_i+\prod_{i=0}^{k-1} G_i V_i
		\right]+\sum_{j=0}^{k-1}\left[
			\prod_{i=1}^{k-1}\rho_i + \left(\dn{G}_j V_j + G_j \dn{V}_j \right) \prod_{\substack{i=0\\i \ne j}}^{k-1} G_i V_i
		\right], 
	\end{equation}
where $\rho_i=\rho\left(\bfx_{i-1}\to\bfx_i\to\bfx_{i+1}\right)$, 
			$G_i = G\left(\bfx_i \lrarrow \bfx_{i+1}\right)$, and \\ 
			$V_i = V\left(\bfx_i \lrarrow \bfx_{i+1}\right)$, 
			for any index $i$.

}

We then compute the normal derivative of our novel term $S_e(\sPath)$ as
\begin{equation}
	\label{eq:diff_Se_twoterms}
	\dnSe(\sPath) = \pfrac{S_e(\sPath)}{t}\dntof(\sPath) + \sum_{i=0,1,k-1,k}\pfrac{S_e(\sPath)}{\bfx_i}\overset{\square}{\bfx_i}.
\end{equation}
\NEW{%
Considering $S_e$ a function of $t$, $\bfx_0$, $\bfx_1$, $\bfx_{k-1}$, and $\bfx_k$, the partial derivatives are evaluated as follows:
	\begin{equation}
		\label{eq:Se_5partials}
		\small
		\begin{split}
			\pfrac{S_e\left(\sPath\right)}{t}=&
			\int_{-\infty}^{\infty}\biggl[{
				\pfrac{L_e}{t}\left(\bfx_0,\bfx_1,t'\right)W_e \left(\bfx_{k-1},\bfx_k,t'+t\right)} \\
			&\quad\quad+L_e\left(\bfx_0,\bfx_1,t'\right)\pfrac{W_e}{t} \left(\bfx_{k-1},\bfx_k,t'+t\right)
			\biggl] \d t'\bigg|_{t=\tof\left(\sPath\right)}, \\
			\pfrac{S_e\left(\sPath\right)}{\bfx_i} =& \left.\int_{-\infty}^{\infty}{
				\pfrac{L_e}{\bfx_i}\left(\bfx_0,\bfx_1,t'\right)W_e \left(\bfx_{k-1},\bfx_k,t'+t\right)
			\d t'}\right|_{t=\tof\left(\sPath\right)}\footnotesize {\text{ for \textit{i}=0,1,}}\\
			\pfrac{S_e\left(\sPath\right)}{\bfx_i} =& \left.\int_{-\infty}^{\infty}{
				L_e\left(\bfx_0,\bfx_1,t'\right)\pfrac{W_e}{\bfx_i} \left(\bfx_{k-1},\bfx_k,t'+t\right)
			\d t'}\right|_{t=\tof\left(\sPath\right)}\footnotesize {\text{ for \textit{i}=\textit{k}-1,\textit{k}.}}
		\end{split}
	\end{equation}
The normal derivative of the path time of flight is 
\begin{align}
	\label{eq:deriv_tof}
		\dntof(\sPath) &= \sum_{i=1}^k \frac{\eta_i}{c} \norm{\bfx_i-\bfx_{i-1}}^\square  +\frac{\dot{\eta_i}}{c}\norm{\bfx_i-\bfx_{i-1}}\\
		&= \sum_{i=1}^k \frac{\eta_{i}}{c} \frac{\left(\bfx_i-\bfx_{i-1}\right)\cdot \overset{\square}{\bfx}_i + \left( \bfx_{i-1}-\bfx_i \right)\cdot\overset{\square}{\bfx_{i-1}}}{\norm{\bfx_i-\bfx_{i-1}}} +\frac{\dot{\eta_i}}{c}\norm{\bfx_i-\bfx_{i-1}} \nonumber.
\end{align}
In both Equations~\eqref{eq:diff_Se_twoterms} and \eqref{eq:deriv_tof} the derivatives of the vertices $\overset{\square}{\bfx_i}$ lie on the velocity field of $\calM$ as $\overset{\square}{\bfx_i}=\mathcal{V}_\calM(\bfx_i)$. 
}

To obtain the final expression for $\dnFtau(\sPath)$, we plug Equation~\eqref{eq:diff_Se_twoterms} into Equation~\eqref{eq:diff_fcalt_twoterms}, which yields:
\begin{equation}
	\label{eq:diff_f_trans}
	\dnFtau(\sPath)=\dnF_s(\sPath) + \mathfrak{T}(\sPath)\, \pfrac{S_e(\sPath)}{t}\, \dntof(\sPath),
\end{equation}
where $\dnF_s(\sPath)$ is defined as
\begin{equation}
	\dnF_s(\sPath) = \dnT(\sPath) {S_e}(\sPath) + \mathfrak{T}(\sPath) \sum_{i=0,1,k-1,k}\pfrac{S_e(\sPath)}{\bfx_i}\overset{\square}{\bfx_i}.
\label{eq:transient_dfn}
\end{equation}
\NEW{The second term in Equation~\eqref{eq:diff_f_trans} models the temporal derivatives for transient rendering (which do not exist in steady-state differentiable rendering), while $\dnF_s(\sPath)$ models the spatial derivatives of the path contribution, similar to the steady-state differential path integral in Equation~\eqref{eq:ss_diff_path_integ}. }

\paragraph{Boundary term}
The main problem in the second term of Equation~\eqref{eq:diff_trans_path_int} (the boundary term) is to determine the domain of integration $\boundary{\Omega}[\ftau]$, in which we will compute the path contribution $\Delta\ftau(\sPath)$,
as well as how this domain evolves with $\btheta$ according to the velocity field $\calV_{\boundary{\calM}}(\sPath)$. 
Recall that, similarly to its steady-state counterpart in Equation~\eqref{eq:ss_diff_path_integ}, $\boundary{\Omega}[\ftau]$ represents the extended boundary of $\ftau$, where we include both the manifold boundaries $\partial\Omega$ and the discontinuity set $\Delta\Omega[\ftau]$ of~$\ftau$ as $\boundary{\Omega}[\ftau]=\partial\Omega\bigcup\Delta\Omega[\ftau]$. For our derivation, we consider the set of discontinuities $\Delta\Omega_k[\ftau]$ for paths of length $k$; it follows that $\Delta\Omega[\ftau]=\bigcup_{k=1}^\infty \Delta\Omega_k[\ftau]$. 
We summarize here the main results of this derivation, and refer the reader to the supplemental material for the full details. 

By the generalized transport theorem, the extended boundary of~$\ftau$ is the union of the boundary of manifold $\Omega_k$, and the discontinuities in the different terms of Equation~\eqref{eq:trans_path_contir_nodelay}, namely the visibility $V$, geometry $G$, scattering $\rho$, and correlated importance $S_e$ terms, as 
\begin{align}
	\label{eq:total_exbd_union}
		\boundary{\Omega}_k[\ftau] =& \partial\Omega_k \cup\Delta\Omega_k[\ftau] \nonumber \\
		=& \partial\Omega_k
		\cup \Delta\Omega_k\left[S_e\right]
		\cup \Delta\Omega_k\left[G_1 \cdots G_k\right] \nonumber \\
		&\cup \Delta\Omega_k\left[V_1 \cdots V_k\right]
		\cup \Delta\Omega_k\left[\rho_1 \cdots \rho_{k-1}\right].
\end{align}
By assuming realistic non-singular scattering functions and light sources following previous work~\cite{li2018differentiable, loubet2019reparameterizing, zhang2020path, bangaru2020unbiased, zhang2019differential}, and applying the product space rule, we can express Equation~\eqref{eq:total_exbd_union} as
\begin{align}
	\label{eq:total_exbd_union_minimal}
	\begin{split}
		\boundary{\Omega}_k[\ftau]
		=& \partial\Omega_k
		\cup \Delta\Omega_k\left[S_e\right] \\
		&\cup\left(\Delta\Omega_k\left[G_1 V_1 \cdots G_k V_k\right]-\partial\Omega_k\right),
	\end{split}
\end{align}
with 
\begin{align}
	\label{eq:disconti_GV_term_path}
		\Delta\Omega_k\left[G_1 V_1 \cdots G_k V_k\right]&-\partial\Omega_k
		= \bigcup_{i=1}^{k} \calM_0 \times\cdots\times\calM_{i-2} \\
		&\times \left( \Delta\calM^2\left[GV\right]-\partial{\calM}^2 \right)
		\times \calM_{i+1}\times\cdots\times\calM_k, \nonumber
\end{align}
where $\Delta\calM^2 [GV]-\partial{\calM}^2$ represents the path segments that intersect a silhouette edge of $\calM$ \NEW{(see Figure~\ref{fig:interior_boundary_sampling} and
Lemma 2.4 in the supplemental)}. Note that under these conditions $\Delta\Omega_k\left[G_1 V_1 \cdots G_k V_k\right]$ represents a similar boundary as in steady state \cite{zhang2020path}, but derived from the generalized transport theorem without restrictions in the dimensionality of the manifolds.

The time-dependent effects on the boundary space $\boundary{\Omega}_k[\ftau]$ are implicitly encoded in $\Delta\Omega_k\left[S_e\right]$. For theoretical Dirac-delta emissions and sensor responses this becomes relevant, while for most practical applications \NEW{(from ultrashort lasers to amplitude-modulated time-of-flight sensors)} this can be handled without additional sampling. Figure~\ref{fig:case_of_temporal_sensitivity} illustrates these practical scenarios.

\subsection{Material Form of Differential Transient Path Integral}
While our differential transient path integral in Equation~\eqref{eq:diff_trans_path_int} is a general formulation, its material form is more efficient for algorithmic implementation. \NEW{Such material form allows integrating over a linear space parametrizing the spatial curved manifold, which in turn simplifies significantly how boundaries are handled~\cite{zhang2020path}. } Thus, instead of integrating over the curved manifold $\calM\left(\btheta\right)$, which lies on $\R^3$, we can integrate over a planar domain $\mathcal{B}\left(\btheta\right)\subset \R^2$, called a reference configuration, when there exists a global parameterization $\hat{\bfx}\left(\cdot,\btheta\right)\colon \mathcal{B}\left(\btheta\right) \to \calM\left(\btheta\right)$. Following Zhang et al.~\shortcite{zhang2020path} we use a barycentric parametrization mapping $\sPath=\sPathMap(\sMatPath,\btheta)$ from $\sMatPath$ to $\sPath$; this results in $\dnmatFtau(\sMatPath)=\dmatFtau(\sMatPath)$, as well as zero local velocity on boundaries and sharp edges. 
This global parameterization converts our differential transient path integral in Equation~\eqref{eq:diff_trans_path_int} into a simpler form:
\begin{equation}
	\label{eq:diff_trans_pi_material}
	\fDiff{I}{\btheta}  = \int_{\hat\Omega}\left(\hat{f}_{\calT}\right)^\cdot(\bar{\mathbf{p}})\, \d \mu(\bar{\mathbf{p}}) 
+ \int_{\partial\hat{\Omega}} \Delta \hat{f}_{\calT}(\bar{\mathbf{p}})  \calV_{\Delta \mathcal{B}}(\bar{\mathbf{p}}) \,\d \mu_{\partial\hat\Omega}(\bar{\mathbf{p}}),
\end{equation}
where $\hat\Omega=\bigcup_{k=1}^\infty \mathcal{B}^{k+1}$ is the space of all material paths $\bar{\mathbf{p}}=\mathbf{p}_0...\mathbf{p}_k$. The terms in Equation~\eqref{eq:diff_trans_pi_material} can be easily defined by considering the equivalent terms in Equation~\eqref{eq:diff_trans_path_int} as functions of material paths $\bar{\mathbf{p}}$ instead of $\sPath$, and taking the Jacobian of  $\hat{\bfx}$ into account, similar to the case of steady-state differentiable rendering~\cite{zhang2020path}. 
In particular, the path contribution in material form is defined as
\begin{equation}
	\label{eq:material_pathcont}
	\hat{\ftau}\left(\bar\bfp\right) =  \hat{\mathfrak{T}}(\bar\bfp)\hat{S_e}(\bfp),
\end{equation}
where $\hat{\mathfrak{T}}\left(\bar\bfp\right)=\mathfrak{T}\left(\hat\bfx(\bar\bfp,\btheta)\right)\prod_{i=1}^{k-1}J(\bfp_i,\btheta)$, with $J(\bfp,\btheta)=\left|\pfrac{\hat{\bfx}(\bfp,\btheta)}{\bfp} \right|$ denoting the Jacobian determinant of the change of variable $\hat\bfx$. The derivatives of the Jacobian determinant $\pfrac{J(\bfp_i,\btheta)}{\btheta}$ are computed in the same fashion as in the steady-state case~\cite{zhang2020path}.

We compute the material form of the correlated importance as
\begin{equation}
	\label{eq:material_correlated}
	\hat{S_e}\left(\bar\bfp\right) = S_e\left(\hat\bfx(\bar\bfp,\btheta)\right)J(\bfp_0,\btheta)J(\bfp_k,\btheta),
\end{equation}
which only depends on the Jacobian of the first and last vertices of the path $\bfp_0$ and $\bfp_k$. 
Finally, the derivative of the correlated importance $\hat{S_e}(\bar\bfp) $ is
\begin{align}
	\label{eq:material_correlated2}
		(\hat{S_e})^\cdot (\sMatPath) & = \pfrac{S_e\left(\hat\bfx(\bar\bfp,\btheta)\right)}{\btheta} J(\bfp_0,\btheta)\,J(\bfp_k,\btheta) \\
		&+ S_e\left(\hat\bfx(\bar\bfp,\btheta)\right) \left[\pfrac{J(\bfp_0,\btheta)}{\btheta}J(\bfp_k,\btheta)
		+ J(\bfp_0,\btheta)\,\pfrac{J(\bfp_k,\btheta)}{\btheta}\right],\nonumber
\end{align}
with $\pfrac{S_e\left(\hat\bfx\left(\bar\bfp,\btheta\right)\right)}{\btheta}$ computed in an analogous way as Equation~\eqref{eq:diff_Se_twoterms}:
\begin{align}
	\label{eq:material_correlated3}
	\pfrac{S_e\left(\hat\bfx(\bar\bfp,\btheta)\right)}{\btheta} = \pfrac{S_e(\hat\bfx(\bfp_i,\btheta))}{t}\pfrac{\tof(\hat\bfx(\bar\bfp,\btheta))}{\btheta} \nonumber\\
	+ \sum_{i=0,1,k-1,k}\pfrac{S_e(\sPath)}{\bfx_i}\pfrac{\hat\bfx\left(\bfp_i,\btheta\right)}{\btheta}.
\end{align}

\begin{figure*}[t]
	\centering
	\includegraphics[width=\columnwidth]{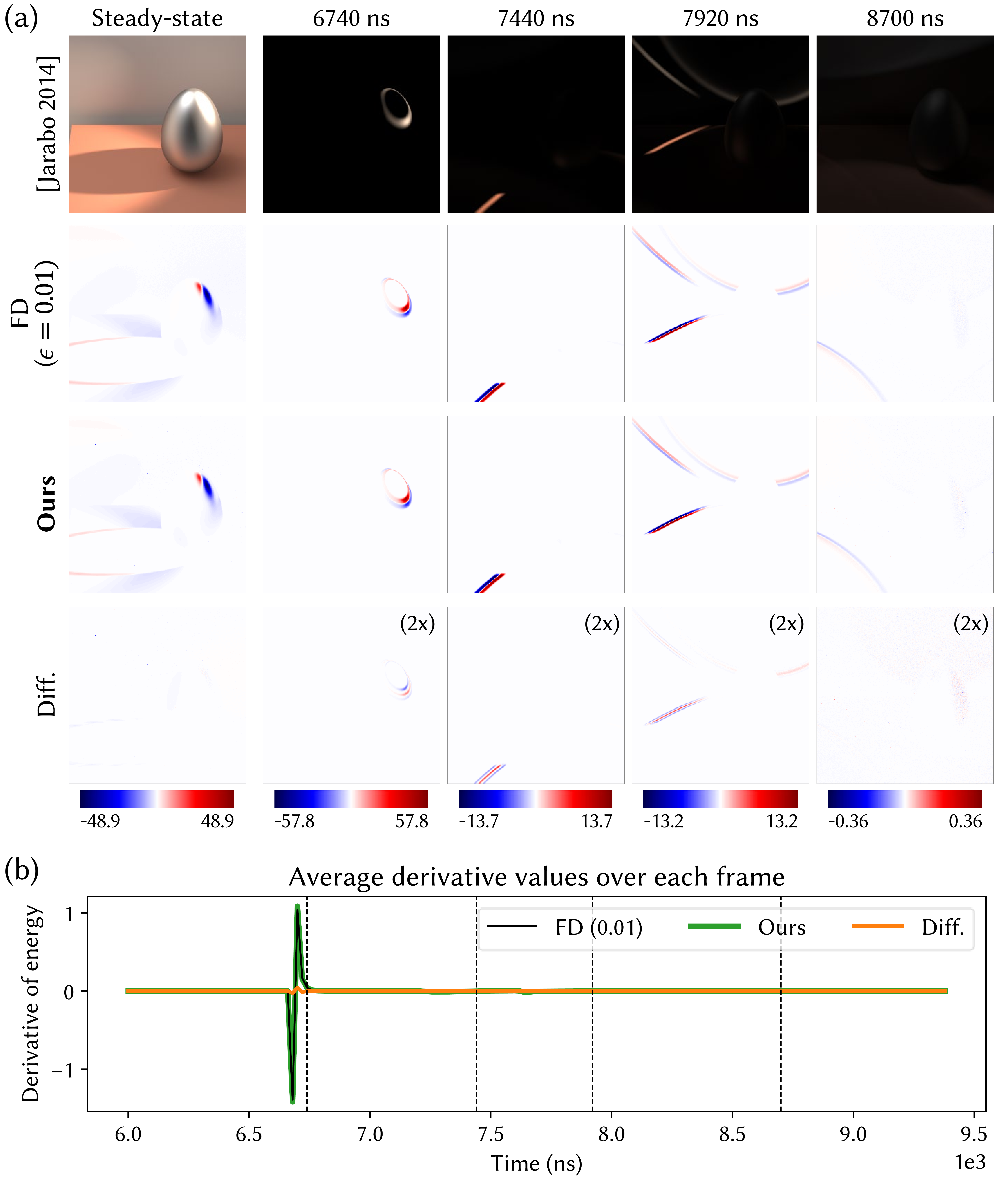} 
	\includegraphics[width=\columnwidth]{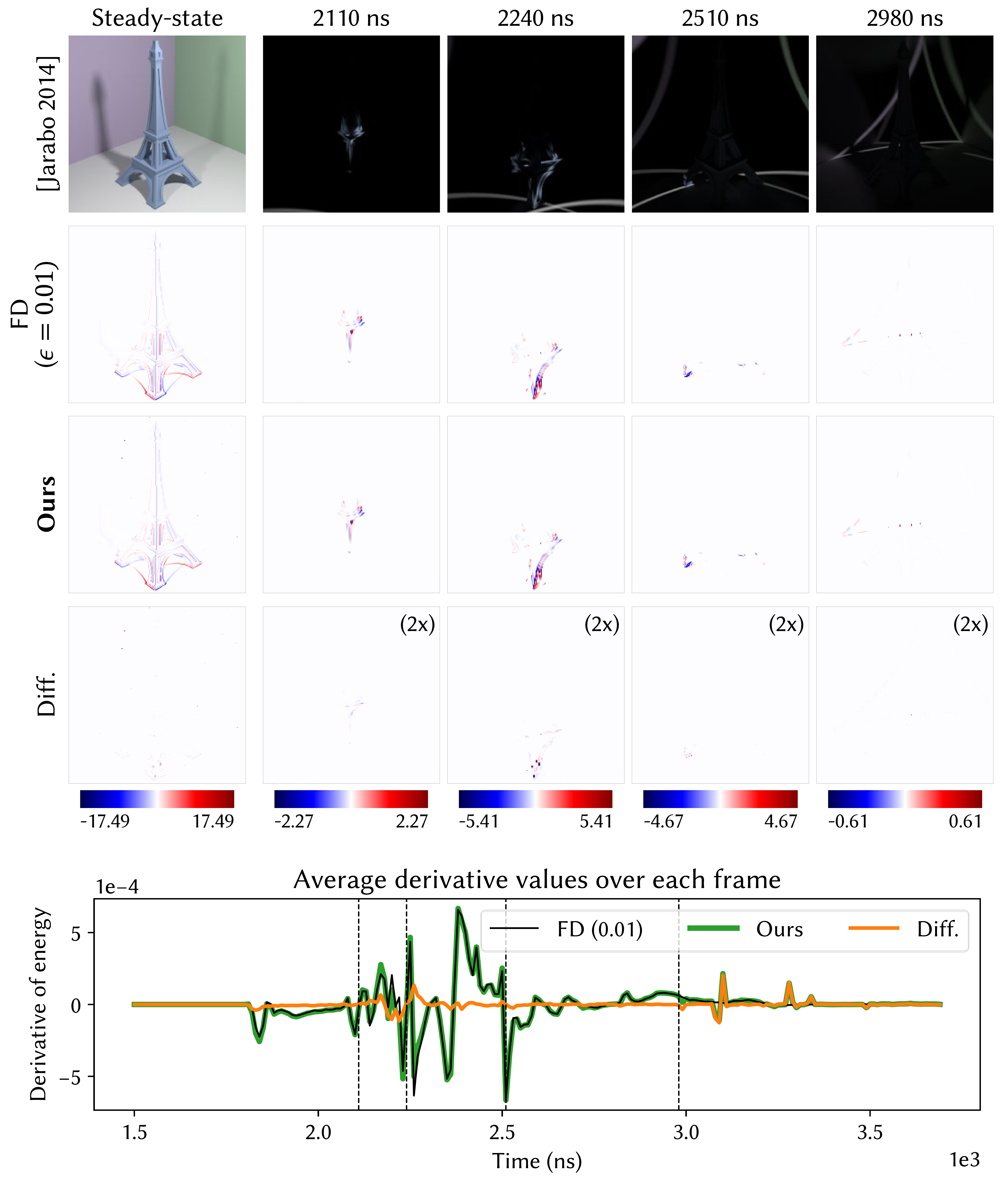}%
	\vspace{-4mm}
	\caption[]{\label{fig:compare_fd}
		Validation of our derivatives against finite differences with $\epsilon=0.01$. Left: \NEW{\textit{Egg}} scene with a moving light source. Right: \NEW{\textit{Tower}} scene with a rotating object. (a)~Transient sequences and derivatives, including the absolute difference amplified by a factor of 2. 
(b)~Gradient plots. \NEW{Dotted vertical lines indicate the timestamp of the frames shown above}. In both scenes our framework matches the finite differences (FD) method closely.
\NEW{The \textit{Egg} scene consists of 1,000 transient frames with $20.0\,\mathrm{ns}$ exposure time. Our result is rendered with \NEWR{410+919} spp (interior + boundary integrals, respectively), which takes $28.9\mathrm{s}$ per frame. The FD result is rendered with \NEWR{4,096} spp at $267.8\mathrm{s}$ per frame. The \textit{Tower} scene consists of 500 transient frames taken with $10.0\,\mathrm{ns}$ exposure time. Our result is rendered with \NEWR{819+1,720} spp at $67.7\mathrm{s}$ per frame. The FD result is rendered with \NEWR{4,096} spp at $144.0\mathrm{s}$ per frame. }}
\vspace{-2mm}
\end{figure*}

\def\bfx{\mathbf x}
\def\bfy{\mathbf y}
\def\bfz{\mathbf z}

\def\bdps{ {\overline{\partial\Omega}} }

\section{Differentiable Monte Carlo Transient Rendering}
\label{sec:ouralgorithm}
We approximate the material form of our differential transient path integral~\eqref{eq:diff_trans_pi_material} as the sum of two Monte Carlo estimators for the interior and boundary integrals, as 
\begin{align}
\fDiff{I}{\btheta} \approx \frac{1}{N_\text{i}} \sum_{j=1}^{N_\text{i}} \frac{\dmatFtau(\sMatPath_j)}{p_\text{i}(\sMatPath_j)} 
 + \frac{1}{N_\text{b}} \sum_{l=1}^{N_\text{b}} \frac{\Delta\matFtau(\sMatPath_l)  \calV_{\boundary{\calB}}(\sMatPath_l)}{p_\text{b}(\sMatPath_l)} ,
\end{align}
where $N_\text{i}$ and $N_\text{b}$ are the number of samples for the interior and boundary integrals, respectively,  and $p_\text{i}(\sMatPath)$ and $p_\text{b}(\sMatPath)$ are the probabilities of sampling the material path $\sMatPath$ for each estimator. 

Similar to previous work on transient rendering~\cite{jarabo2014framework}, we reconstruct the whole temporal domain at the same time by reusing samples between different time frames. We track the time of flight of the path $\sPathMap(\sMatPath_j,\btheta)$, and bin the path's contribution along the temporal domain. The temporal footprint of the path $\sPathMap(\sMatPath_j,\btheta)$ is selected based on a time window determined by the non-zero regions of $S_e\left(\sPathMap(\sMatPath_j,\btheta)\right)$\NEW{, which acts as a temporal smoothing kernel potentially removing temporal discontinuities between frames (see Figure~\ref{fig:case_of_temporal_sensitivity}).  }

\paragraph{Interior integral estimator}
We estimate the interior integral by tracing the path $\sPathMap(\sMatPath_j,\btheta)$ and computing $\dmatFtau(\sMatPath_j)$. Different from Zhang et al.~\shortcite{zhang2020path}, we track the temporal delay of the path encoded in $\tof(\sPathMap(\sMatPath_j,\btheta))$. Note that during this process we compute the throughput $\hat{\mathfrak{T}}(\sMatPath_j)$ and its derivative $\left(\hat{\mathfrak{T}}\right)^\cdot(\sMatPath_j)$, which are later convolved by $\left(\hat{S_e}\right)^\cdot(\sMatPath_j)$ and $\hat{S_e}(\sMatPath_j)$ when binning the contribution on the temporal domain following Equation~\eqref{eq:diff_fcalt_twoterms}.

\paragraph{Boundary integral estimator}
The main difference with respect to the steady-state boundary integral is the existence of the discontinuity term $\Delta\Omega_k[S_e]$. 
\NEW{For most practical situations, both light sources and sensors have a finite temporal response making $S_e\left(\sPathMap(\sMatPath_j,\btheta)\right)$  continuous (Figure~\ref{fig:case_of_temporal_sensitivity}), so $\Delta\Omega_k[S_e]$ vanishes}; the transient boundary integral estimation is then similar to its steady-state counterpart, except that we again need to keep track of the path time of flight $\tof(\sPathMap(\sMatPath_l,\btheta))$ for binning.

\begin{figure*}[thpb]
	\centering
	\includegraphics[width=\columnwidth]{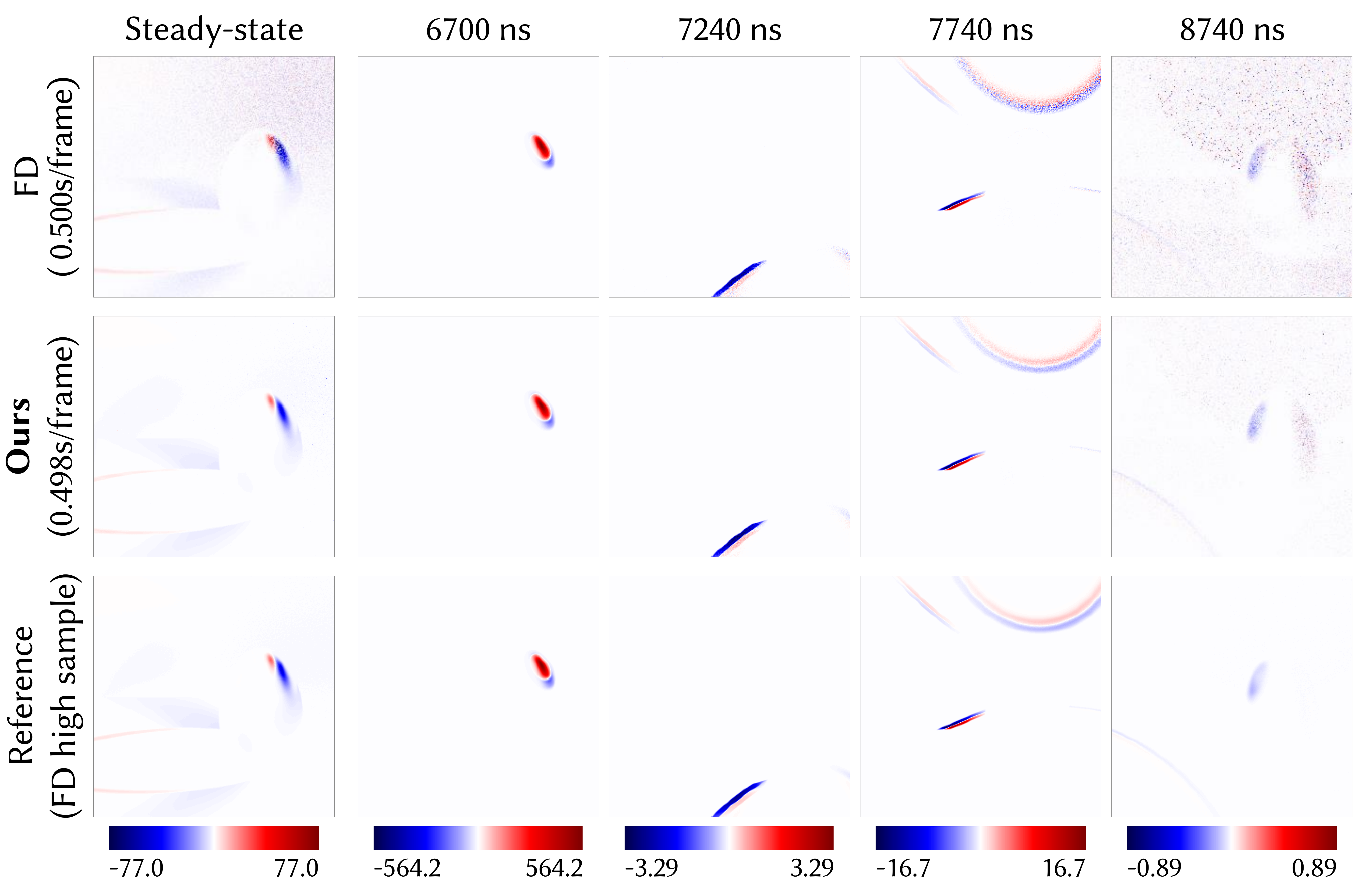} 
	\includegraphics[width=\columnwidth]{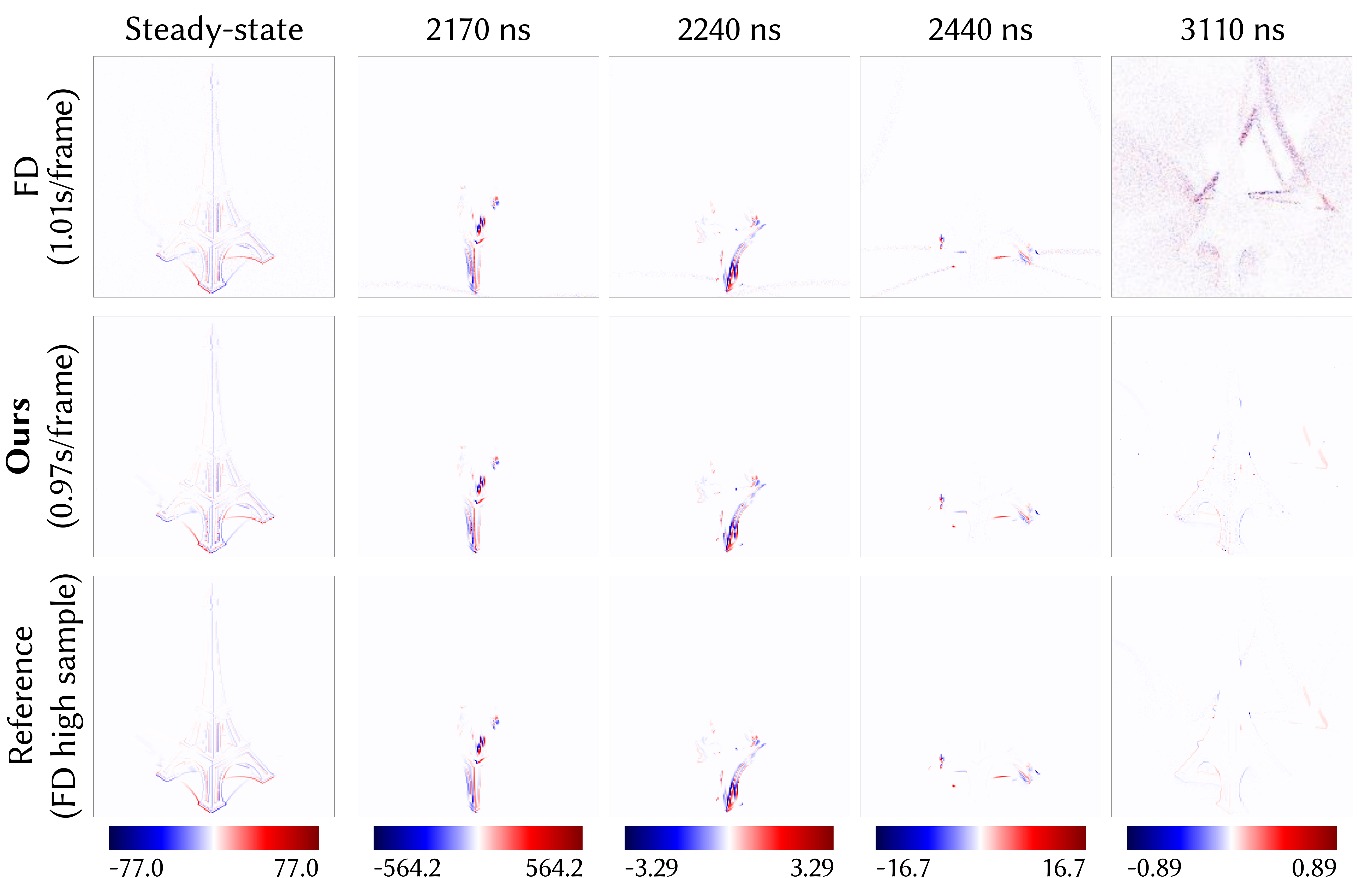}%
	\vspace{-3mm}
	\caption[]{\label{fig:compare_fd_speed}
		\NEW{Equal-time comparisons for FD and our derivatives on the \textit{Egg} (left) and \textit{Tower} (right) scenes. As the first and second rows show, our derivatives are visibly less noisy. The third row shows a high-sample reference (4,096 spp) rendered with FD, similar to the second row of Figure~\ref{fig:compare_fd}.}}
		\vspace{-2mm}
\end{figure*}
\begin{figure}[htpb]
	\centering
	\includegraphics[width=\columnwidth]{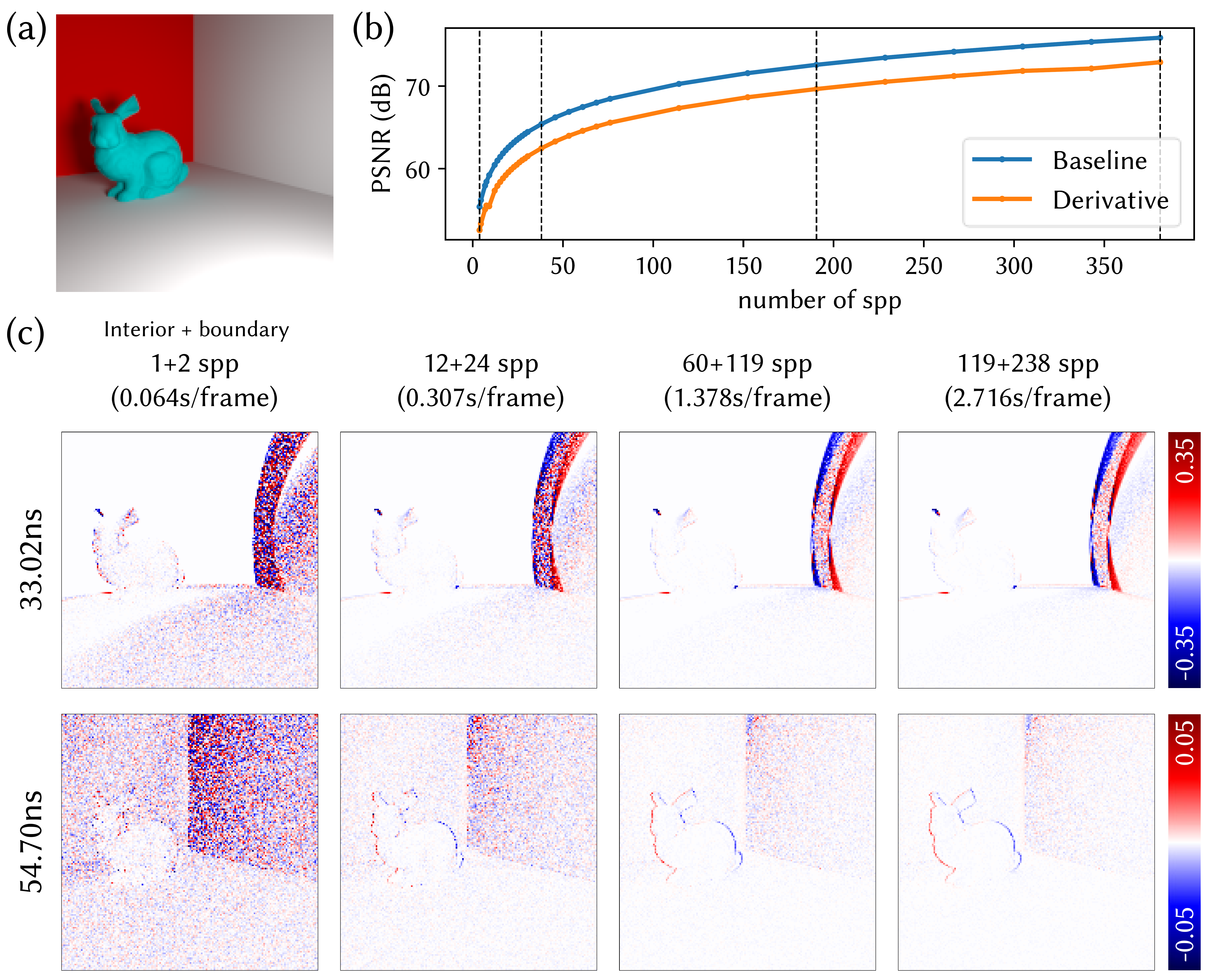}%
	\vspace{-3mm}
	\caption[]{\label{fig:bunny-variance}
		\NEW{Evolution of variance with the number of samples per pixel. (a) Steady-state scene; (b) PSNR of our baseline~\cite{jarabo2014framework} and derivative transient images;  (c) selected derivatives of transient frames, indicated by the dashed lines in (b) for each spp (interior plus boundary integrals). The scene parameter being optimized is the $x$-translation of the camera.}}
		\vspace{-2mm}		
\end{figure}

\section{Results and Applications}
\label{sec:results-applications}

We have implemented our differentiable transient rendering on top of the code provided by Zhang et al.~\shortcite{zhang2020path}, by extending the bidirectional path tracer to keep track of the propagation delay of the paths.  
\NEW{We compute the interior integral using automatic differentiation on top of next-event estimation and radiance-based multiple importance sampling. 
For the boundary integral, we use the multidirectional sampling of boundaries, next-event estimation, and the grid-based importance sampling proposed by Zhang et al.}
For temporal reconstruction we use histogram binning, though more advanced density estimation techniques~\cite{jarabo2014framework} could be used without changes in our derivatives estimation. \NEW{We refer to the supplemental material for pseudocode describing the computation of both the interior and boundary integrals.} 

All our results have been obtained using a conventional desktop computer equipped with an Intel Core I9-10920X CPU of 3.5\,GHz with 128\,GB RAM, and an NVIDIA Titan RTX graphics card. Without loss of generality, the examples shown are based on a triangular correlated importance function $S_e$ (using discontinuous $L_e$ and $W_e$ functions). Please refer also to our supplementary video for transient rendering results.

We first compare our derivatives against the finite differences (FD) method on two different scenes. \NEW{To obtain the FD, we use our transient renderer by computing radiance with the same (radiance-based) sampling routines we use for the interior integral. }
Figure~\ref{fig:compare_fd}, left, shows the \textit{Egg} scene, with a vertical translation of the light source as the varying parameter. 
Figure~\ref{fig:compare_fd}, right, shows the \textit{Tower} scene, where the tower rotates along its vertical axis. 
In both scenes our method matches the results from finite differences closely. 
\NEW{In addition, Figure~\ref{fig:compare_fd_speed} provides equal-time comparisons, showing how our method provides less noisy results than FD.}

\NEW{We also evaluate  in Figure~\ref{fig:bunny-variance} the evolution of variance with the number of samples per pixel.
We observe that it follows a trend similar to baseline  transient rendering~\cite{jarabo2014framework}.}

\begin{figure}[htbp]
	\centering
	\includegraphics[width=\columnwidth]{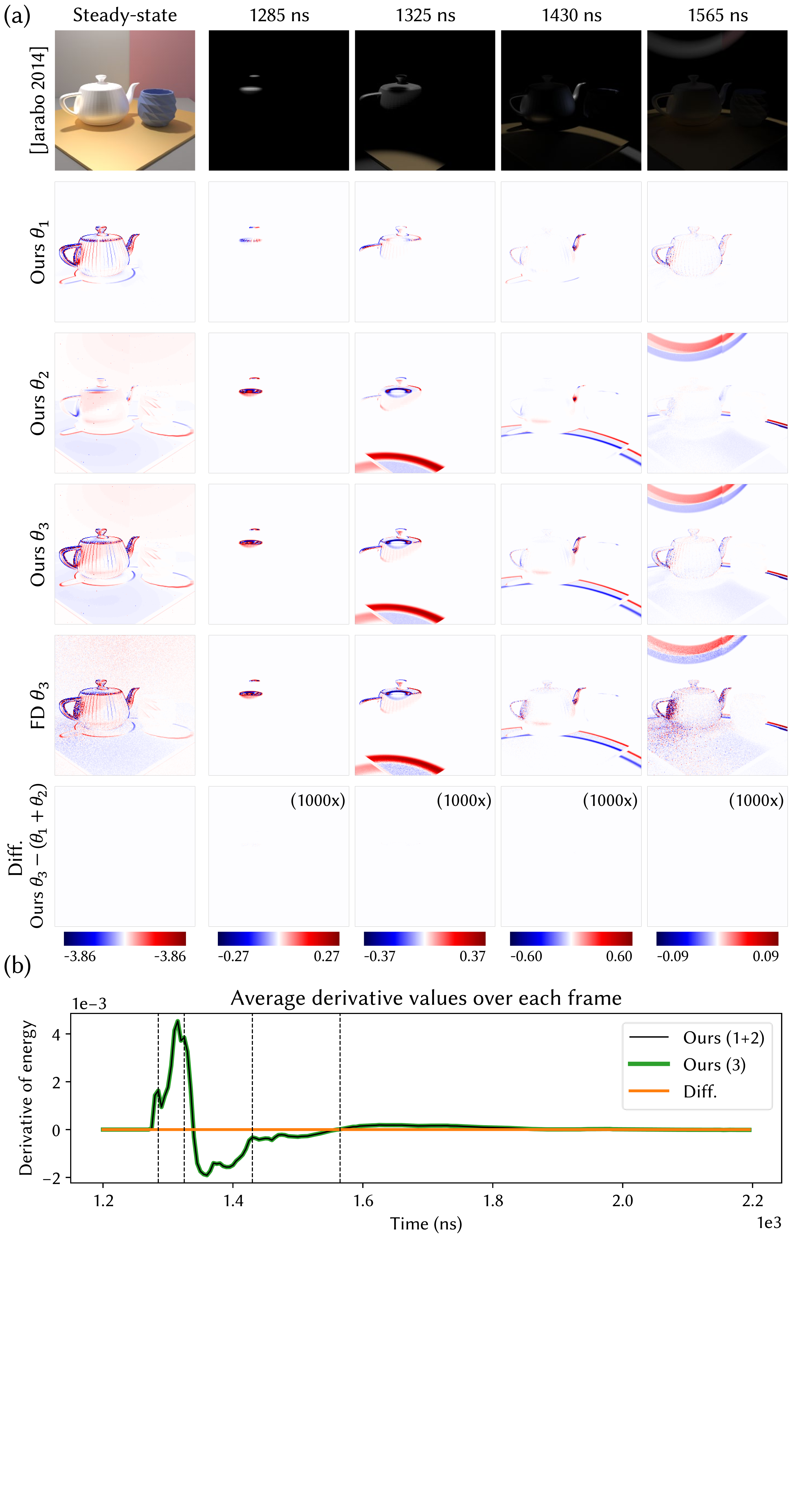}%
	\vspace{-4mm}
	\caption[]{\label{fig:compare_fd3}
Validation of our derivatives when varying several parameters at the same time. a) Transient sequence and derivatives. $\theta_1$ corresponds to the translation of the teapot, $\theta_2$  to the translation of the light source, and $\theta_3$  to the combined translation of both teapot and light source. The absolute difference (x1000) shows how the sum of scene derivatives w.r.t. $\theta_1$ and $\theta_2$ is identical to the scene derivatives w.r.t. $\theta_3$. 
(b) Radiance and absolute difference plots. Dotted vertical lines indicate the exact frames shown above. \NEW{The scene consists of 600 transient frames with an exposure of $5.0\,\mathrm{ns}$. Our results are rendered with \NEWR{683+1,382} spp (interior + boundary integrals, resp.) taking $94.7\mathrm{s}$ per frame. The FD result is rendered with \NEWR{$4\times683$} spp taking $173.4\mathrm{s}$ per frame. }}
\vspace{-4mm}
\end{figure}

We have also validated our scene derivatives when varying several parameters at the same time. 
 Figure~\ref{fig:compare_fd3} shows the transient sequence for the \textit{Teapot} and its scene derivatives  with respect to three parameters.  $\theta_1$ refers to a translation of the teapot, $\theta_2$ to a translation of the light source, and $\theta_3$ to the combined motion of the object and the light source. As the absolute difference row shows (amplified by a factor of 1000) $\pfrac{I}{\theta_1}+\pfrac{I}{\theta_2}=\pfrac{I}{\theta_3}$ when $\theta_3$ is the combined motion of $\theta_1$ and $\theta_2$, which is a general key property of differentiable rendering.

\begin{figure}[pt]
	\vspace{3mm}
	\centering
	\includegraphics[width=\columnwidth]{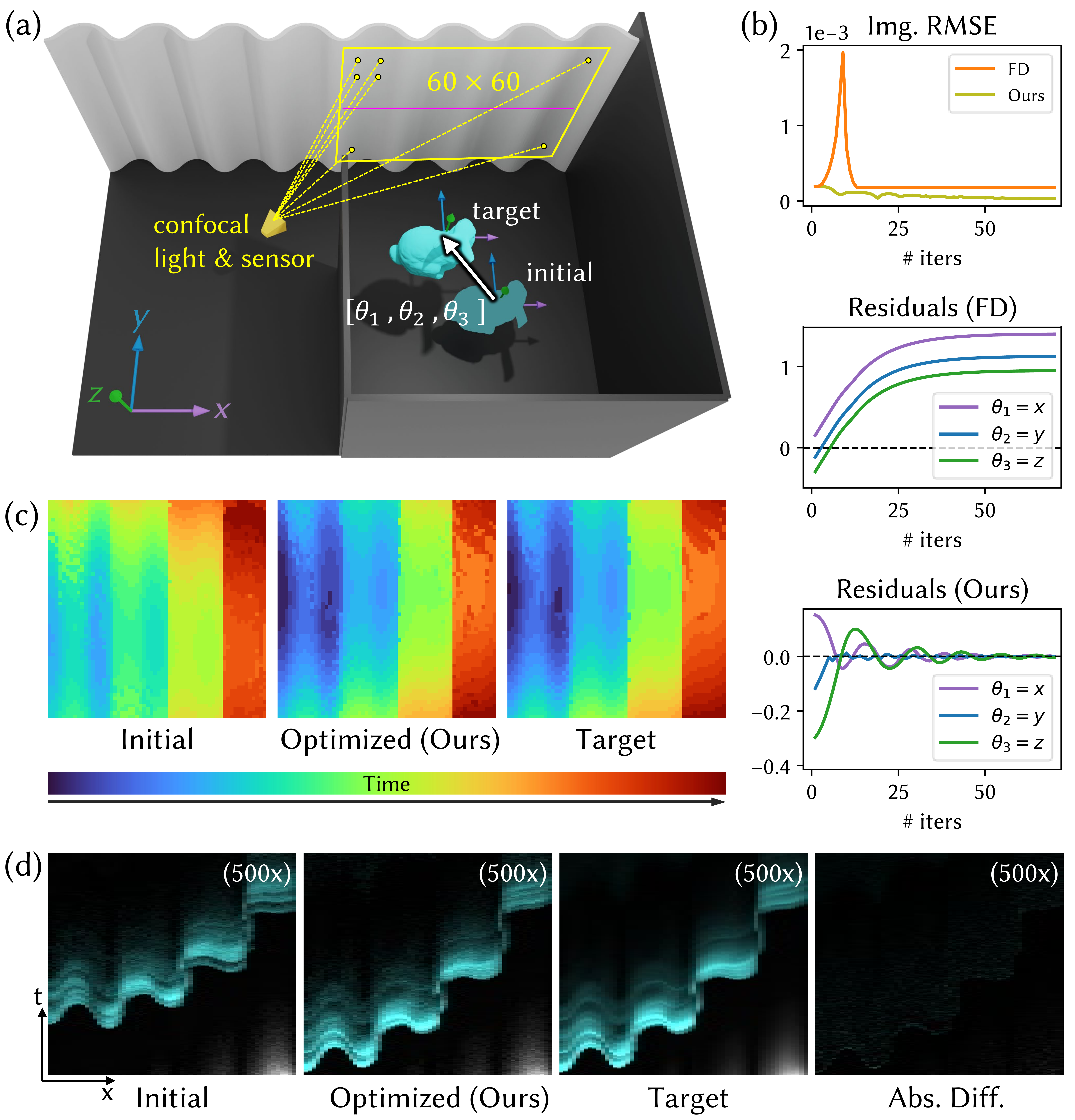}%
	\vspace{-3mm}
	\caption[]{\label{fig:curved-nlos-imaging}
	Three-parameter optimization in an NLOS tracking scenario with a \textit{wavy} relay wall. (a) The hidden bunny moves to the target position, which implies changes in three parameters: $\theta_1=x$, $\theta_2=y$, and $\theta_3=z$. The yellows dots illustrate sampled positions.
	\NEW{(b) RMSE of rendered transient images (top), as well as residuals for each parameter obtained by FD (middle) and our method (bottom).}
	(c) Color-coded transient results: initial, optimized parameters after 69 iterations \NEW{using our method}, and target. (d) Streak images in $x-t$  for the coordinates indicated by the magenta line in (a).  }
	\vspace{5mm}
	\centering
	\includegraphics[width=\columnwidth]{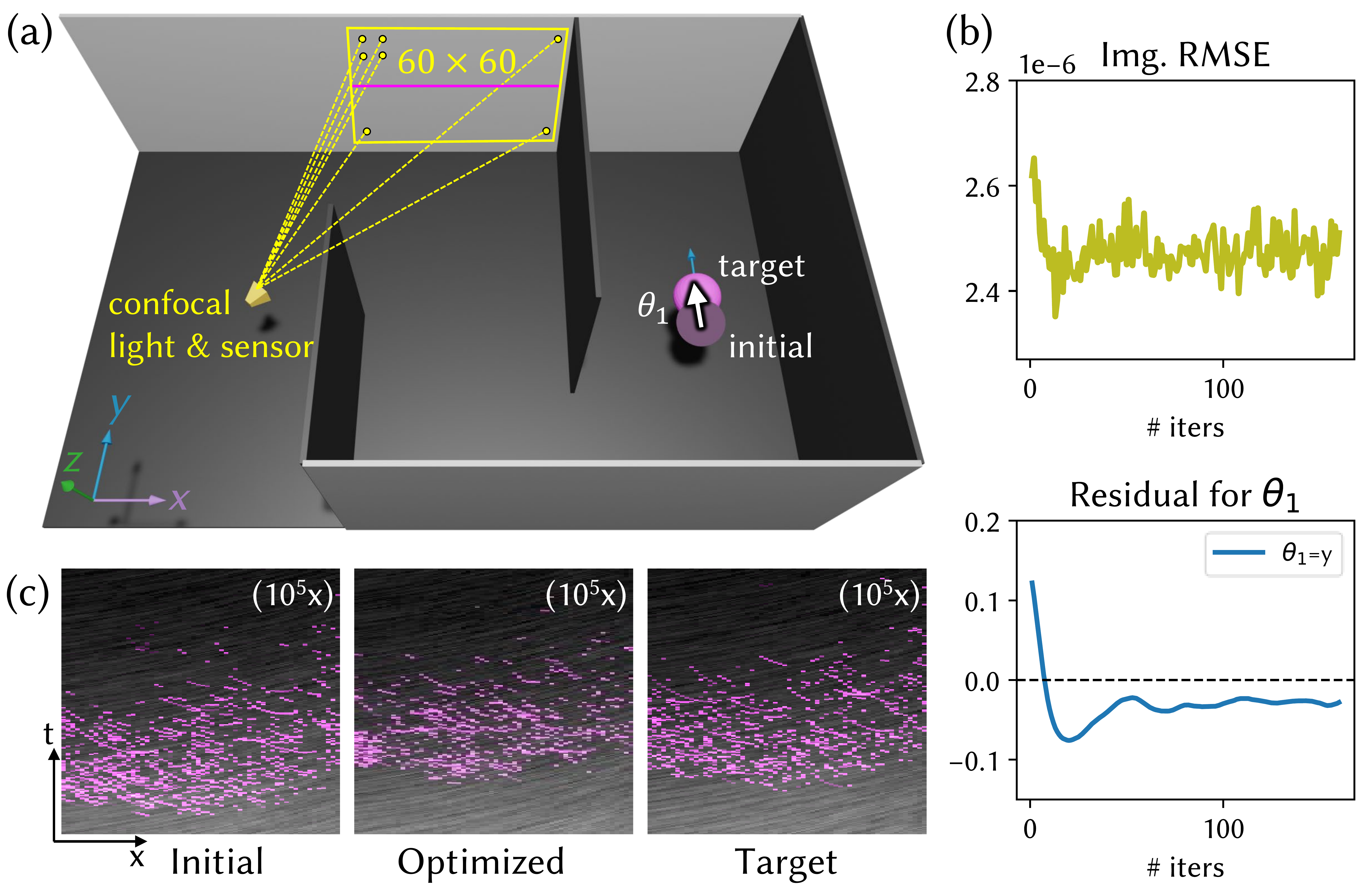}%
	\vspace{-3mm}
	\caption[]{\label{fig:nlos-tracking}
NLOS tracking around two corners. (a) Scene setup, where the hidden object moves along one axis. 
(b) RMSE of the image and residual of the optimized parameter. (c) 
Streak images in $x-t$  for the coordinates indicated by the magenta line in (a) for the initial situation, optimized parameter after 159 iterations, and target. \NEW{Even in this challenging scenario, our method converges with just a small residual bias.} }
\end{figure}

\subsection{Applications}
\label{sec:applications}

\begin{figure*}[htbp]
	\centering
	\includegraphics[width=0.47\linewidth]{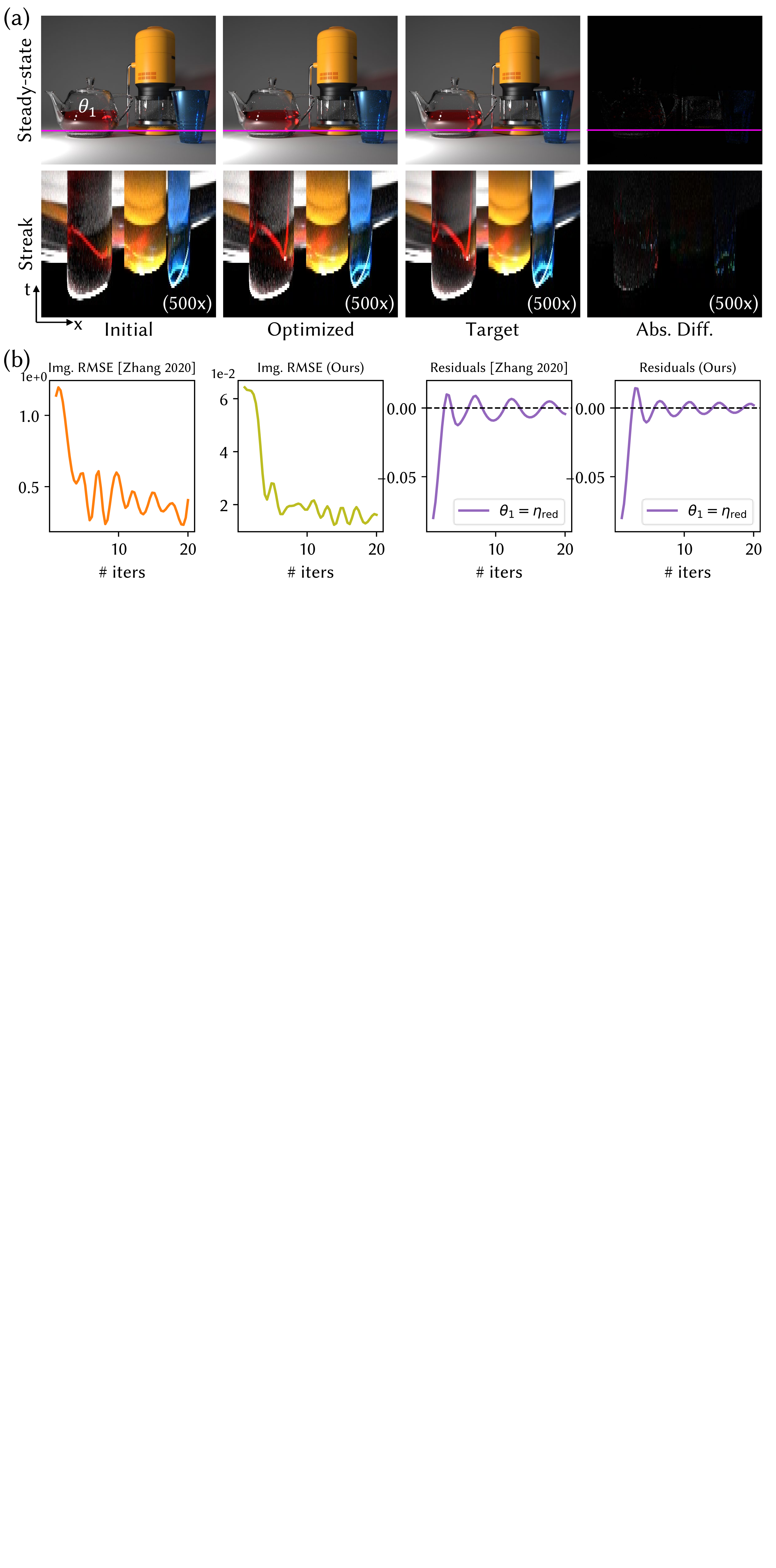}%
	\includegraphics[width=0.47\linewidth]{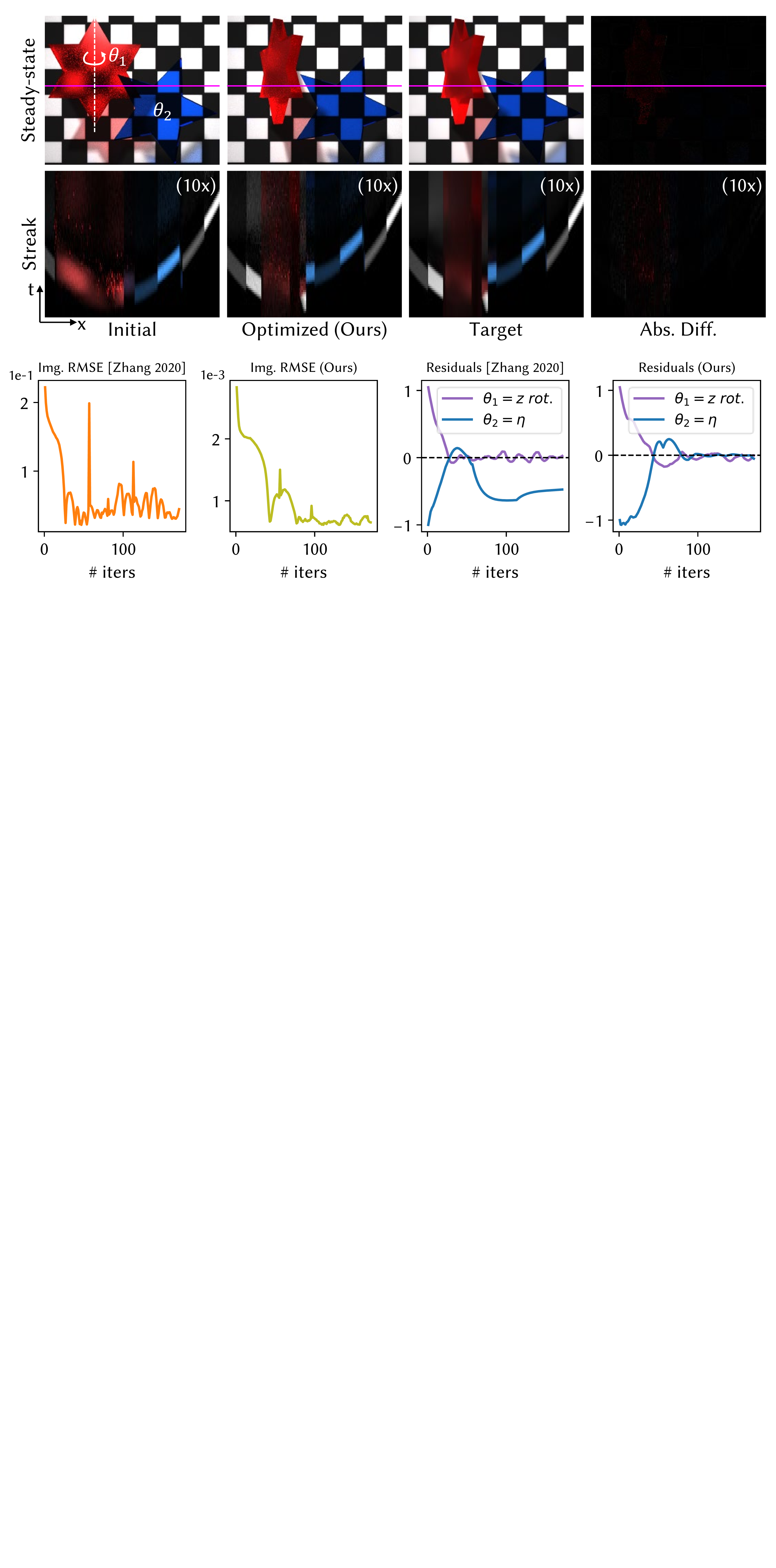}%
	\vspace{-4mm}
	\caption[]{\label{fig:transparent-rendering}
Two examples of parameter optimization involving indices of refraction. 
Left: The same \textit{Coffee} scene as in Figure~\ref{fig:teaser}, depicting a glass teapot with red tea. We optimize the index of refraction of the tea $\theta_1 $ from an arbitrary value (1.33) to the target value (1.41).
Right: \textit{Stars} scene showing two translucent glass stars.
$\theta_1$ represents the single-axis rotation of the red star while $\theta_2$ represents the index of refraction of the blue star, changing from 1.5 to 2.5.
For both scenes:  (a) The top row shows steady-state images for the initial, optimized, and target parameters, as well as the difference between the last two. The bottom row shows the respective $x-t$ streak images, corresponding to the magenta line. 
It can be observed how changing the index of refraction affects not only the BSDF of the object, but also the time of flight of light. 
(b) RMSE of the image and evolution of the residuals for each parameter,
\NEW{as well as equal-sample comparison of our results with Zhang et al.'s steady-state renderer~\shortcite{zhang2020path}. 
	Although the error scales are not directly comparable, the optimization of the index of refraction in the steady-state approach converges more slowly in the \textit{Coffee} scene, while failing to converge in \textit{Stars}.
}}
\end{figure*}

We begin by presenting the first demonstrations of two challenging NLOS scenarios where the time of flight of light plays a key role. 
We use ADAM optimizer with a scene-dependent learning rate
(see Table~\ref{tb:opt_info}). 
First, our \textit{Wavy} scene in Figure~\ref{fig:curved-nlos-imaging} shows  a variation of the classic confocal NLOS imaging setup; \NEW{we lift the requirement of using a planar surface as relay wall, and use a wavy surface instead.} In the hidden scene, the cyan bunny moves to a different target position. We jointly optimize the three coordinates $x$, $y$, and $z$ defining its position. Our sampling resolution is $60\times 60\times 1200$.
\NEW{We include a comparison of our derivative computation and FD equipped with the same optimizer and learning rate.
Our results are rendered with \NEWR{$17$ and $34$}\,spp for interior and boundary integrals, respectively, while FD results are rendered with \NEWR{$4\times17$}\,spp. 
FD takes $61$\,min.~per iteration, compared to $36$\,min.~in our approach.} 
\NEWRR{Moreover, it can be seen how FD fails to optimize the parameters successfully. The oscillating nature of converge in our method is similar to a general stochastic gradient descent.}

Next, we show a particularly difficult NLOS scenario: being able to track an object around \textit{two} corners. Our \textit{Corners }scene can be seen in Figure~\ref{fig:nlos-tracking}, where the hidden object moves along one axis. 
Despite the extremely challenging conditions of the scene, we are able to provide reasonable optimized results.

Figure~\ref{fig:transparent-rendering} shows two different cases of parameter optimization.
The \textit{Coffee} scene  (also shown in Figure~\ref{fig:teaser}) includes two transparent, colored objects: a glass teapot with reddish tea and an empty blue cup.
Using our scene derivatives,
we optimize the \NEW{index of refraction of the tea in the teapot}
from \NEW{an} arbitrary initial value to the final target value.
Last, the \textit{Stars} scene is composed of two translucent, colored stars. The red star rotates around one axis, while the index of refraction of the blue star changes. In both scenes, the streak images show how changing the index of refraction affects the time of flight of light through the transparent objects. 

\NEW{Figure~\ref{fig:transparent-rendering} also includes and equal-sample
comparison with Zhang et al.'s steady-state renderer~\shortcite{zhang2020path}. It can be seen how the index of refraction converges faster with our method in the  \textit{Coffee} scene, while for \textit{Stars} the steady-state approach fails to converge. }
Table~\ref{tb:opt_info} summarizes the main information for all scenes shown. Please refer to the supplemental video for the full transient sequences.

\begin{table}[h]
	\caption{\label{tb:opt_info} 
		\NEW{Rendering and} optimization data for our application scenarios. \NEW{Since there is no geometry change in the \textit{Coffee} scene, we do not need to perform boundary integral for this scene. Note that \textit{Wavy} and \textit{Corners} scenes requires high spps due to indirect-dominant NLOS scenario.}
	}
\small
	\vspace{-3mm}
	\begin{tabular}{ccccc}
		\thickhline
		\textbf{Scene} & \textbf{Coffee} & \textbf{Stars}& \textbf{Wavy} & \textbf{Corners} \\
		\hline
		\NEW{\# spp (interior)} & \NEWR{$41$} & \NEWR{$1$} & \NEWR{$17$} & \NEWR{$85$} \\
		\NEW{\# spp (boundary)} & \NEW{$0$} & \NEWR{$4$} & \NEWR{$34$} & \NEWR{$85$} \\
		\NEWR{\# frames} & \NEWR{$250$} & \NEWR{$800$} & \NEWR{$1200$} & \NEWR{$1200$} \\
		\# param. & $2$ & $2$ & $3$ & $1$ \\
		\# iter. & 19 & $129$ & $69$ & 159  \\
		Time per iter. & $28.5\mathrm{min.}$ & $5.7\mathrm{min.}$ & $36\mathrm{min.}$ & $32.8\mathrm{min.}$  \\
		Learning rate & 0.07 & $0.07$& $0.07$ & $40.0$ \\
		\thickhline
	\end{tabular}
		\vspace{-3mm}
\end{table}

\section{Discussion and Conclusions}
\label{sec:discussion}
We have presented a framework for differential transient rendering, based on the key observation that in transient state scattering events at path vertices are no longer independent. Instead, tracking the time of flight of light requires treating such scattering events jointly as a multidimensional, evolving manifold. For this, we have relied on the generalized transport theorem, and introduced a  correlated importance function which allows us to handle discontinuous importance functions for both the light and the sensor. We have shown how to incorporate our framework in a Monte Carlo renderer, and demonstrated its application in several challenging scenarios. 
\NEW{The cost of computing our derivates depends on the scene, due to next-event estimation and importance sampling over edges in the boundary integral (similar to Zhang's method). In the worst case (\textit{Tower}) the overhead is 72\% with respect to standard transient rendering, whereas for our NLOS results (\textit{Wavy} and \textit{Corners}) it is just ~10\%. Since a finite differences approach scales with the number of optimized parameters, our method is faster for all scenes shown.}

\paragraph{Relation to the steady-state differential path integral}
Our work can be seen as a generalization of the steady-state differential path integral~\cite{zhang2020path} to higher-dimensional domains, and handling discontinuities beyond the geometric visibility term. As expected, when our time-dependent terms disappear, the temporal boundary in path space vanishes and our differential transient path integral converges to its steady-state counterpart.

\paragraph{Limitations and future work}
Several exciting avenues of future work lie ahead. From a theoretical point of view, it would be interesting to include Dirac delta functions in the temporal domain for both the light and the sensor.  \NEW{From a more practical perspective, assuming non-negligible scattering delays would allow handling fluorescent materials~\cite{gutierrez2008visualizing}, thus extending the applicability of our method to time-resolved fluorescence lifetime imaging (FLI)~\cite{satat2015locating}. This would require re-introducing the temporal dimension on the paths throughput, as well as the scattering delays inside the path time of flight. }
Extending our differentiable transient framework to participating media~\cite{zhang2021path} should be possible, adding \textit{indirect shadow vertices} to control the path duration during sampling~\cite{jarabo2014framework}. Exploiting this additional degree of freedom is a promising avenue for improving sampling of the derivatives. 
Currently our work shares Zhang et al.'s~\shortcite{zhang2020path} limitation of lack of reciprocity.
However, by using the generalized manifold transport theorem over the $k$-dimensional evolving manifold of each path, a fully reciprocal boundary path space $\bdps$ and normal velocity $\mathcal{V}_{\bdps}$ could be derived. 
\NEW{Last, our framework can be used to optimize any parameters describing the scene, including geometry, light sources, or materials. However, similar to Zhang's work \shortcite{zhang2020path}, our current implementation is memory-bounded, which in our case is aggravated by having to compute a full transient sequence.
Computing the derivatives in more complex scenarios with potentially thousands of parameters would benefit from faster differentiation strategies, which could be obtained by combining our work with recent advances on efficient backpropagation of derivatives~\cite{nimierdavid2020radiative,vicini2021path}.}

\section*{Acknowledgments}
We want to thank other lab members for proofreading and the reviewers for their feedback. 
M. H. Kim acknowledges that this work is supported by Samsung Research Funding \& Incubation Center of Samsung Electronics (SRFC-IT2001-04).
D. Gutierrez and A. Jarabo acknowledge additional support by the European Research Council (ERC) under the EU Horizon 2020 research and innovation program (project CHAMELEON, grant No 682080), the EU MSCA-ITN program (project PRIME, grant No 956585) and the Spanish Ministry of Science and Innovation (project PID2019-105004GB-I00).

\bibliographystyle{ACM-Reference-Format}
\bibliography{bibliography}

\appendix

\section{Algorithm tables}
We append Algorithms~\ref{alg:interior} and \ref{alg:boundary} for estimating the interior and boundary integrals, respectively. Variables $d$ and $\dot{d}$ store the optical path length and its derivative with respect to the scene parameter, respectively. In both algorithms, $S_e.\mathrm{range}\left(d\right)$ returns the list of indices of temporal bins in which the sampled path is stored. In Algorithm~\ref{alg:boundary}, $\mathrm{EstimateSensorSubpath}$ and $\mathrm{EstimateSourceSubpath}$ can be performed as in Algorithm~\ref{alg:interior}, without computing the derivative terms $\dot{f}$ and $\dot{d}$.

\begin{algorithm}[htpb]
	\caption{Estimating the interior integral}\label{alg:interior}
	\SetAlgoLined
	\KwData{scene, pixel index $\left(i,j\right)$, max bounce $k$}
	\KwResult{Rendered temporal histogram of the $\left(i,j\right)$-th pixel $I\left[i,j,\cdot\right]$ and its scene derivative $\dot{I}\left[i,j,\cdot\right]$}
	
	$\bfx\left[k+2\right],\bfy\left[k+1\right] \leftarrow$ New arrays of 3D positions on the scene geometry\;
	$d_\bfx \left[k+2\right],\dot{d}_\bfx\left[k+2\right] \leftarrow$ New arrays of floating numbers (path distance)\;
	$d_\bfy\left[k+1\right],\dot{d}_\bfy \left[k+1\right] \leftarrow$ New arrays of floating numbers (path distance)\;
	$f_\bfx \left[k+2\right],\dot{f}_\bfx\left[k+2\right] \leftarrow$ New arrays of floating numbers (throughput)\;
	$f_\bfy \left[k+1\right],\dot{f}_\bfy\left[k+1\right] \leftarrow$ New arrays of floating numbers (throughput)\;
	
	$\bfx\left[0\right]\leftarrow$ camera position, $\left(f_\bfx\left[0\right],\dot{f}_\bfx\left[0\right]\right)\leftarrow\left(1,0\right)$\;
	
	\For{$1\le i < k+2$}{
		\eIf{$i=1$}{
			Sample $\left(\homega_o,p\right) \sim \mathbb{P}_{\text{cameraPrimiryRay},ij}$\;
			
		}{
			Sample $\left(\homega_o,p\right)\sim\mathbb{P}_\text{brdf}\left(\bfx\left[i-1\right],\homega_i,\cdot\right)$\;
		}
		$\bfx_\text{temp}\leftarrow \text{rayTrace}\left(\bfx\left[i-1\right],\homega_o\right)$\;
		\eIf{$\bfx_\text{temp}$ is valid}{
			$\bfx\left[i\right]\leftarrow\bfx_\text{temp}$\;
			$\left(\alpha,\dot \alpha\right)\leftarrow$ The value and scene derivative of:
			$\rho\left(\bfx\left[i-2\right]\to\bfx\left[i-1\right]\to\bfx\left[i\right]\right)G\left(\bfx\left[i-1\right],\bfx\left[i\right]\right)J\left(\bfx\left[i\right]\right)$\;
			$p\leftarrow p\left| \hat{n}\left(\bfx\left[i\right]\right)\cdot -\homega_o \right| / \norm{\bfx\left[i\right]-\bfx\left[i-1\right]}^2$\;
			$f_\bfx\left[i\right]\leftarrow f_\bfx\left[i-1\right]\alpha / p$\;
			$\dot{f}_\bfx\left[i\right]\leftarrow \left(\dot{f}_\bfx\left[i-1\right]\alpha + f_\bfx\left[i-1\right]\dot\alpha\right) / p$\;
			$\left(\delta,\dot \delta\right)\leftarrow$ The value and scene derivative of:
			$\eta\norm{\bfx\left[i\right]-\bfx\left[i-1\right]}$\;
			$d_\bfx\left[i\right]\leftarrow d_\bfx\left[i-1\right]+\delta$\;
			$\dot{d}_\bfx\left[i\right]\leftarrow \dot{d}_\bfx\left[i-1\right]+\dot\delta$\;
			$\homega_i \leftarrow -\homega_o$\;
		}{
			break\;
		}
		
	}
	
	Sample $\bfy\left[0\right]\sim\mathbb{P}_\text{emitter}$, $f_\bfy\left[0\right]\leftarrow 1/\mathbb{P}_\text{emitter}\left(\bfy\left[0\right]\right)$\;
	Similarly construct the light subpath $\bfy\left[\right], d_\bfy\left[\right], \dot{d}_\bfy\left[\right], f_\bfy\left[\right], \dot{f}_\bfy\left[\right]$.\;
	
	\For{$0\le s \le k-1$}{
		$\left(f,\dot f\right) \leftarrow \text{combineSubpaths}\left(\bfx\left[0:s+1\right], \bfy\left[0:k-s\right]\right)$\;
		$\left(\delta,\dot{\delta}\right)\leftarrow$ The value and scene derivative of:
		$\eta\norm{\bfx\left[s\right]-\bfy\left[k-s-1\right]}$\;
		$d\leftarrow d_\bfx\left[s\right]+d_\bfy\left[k-s-1\right]+\delta$\;
		$\dot d\leftarrow \dot {d}_\bfx\left[s\right]+\dot{d}_\bfy\left[k-s-1\right]+\dot{\delta}$\;
		$w \leftarrow \mathrm{CombinationStrategy}$;\ \ \ \ \ // Use Chapter 9 in \cite{veach1997robust}\\
		\For{$l\in S_{e}.\mathrm{range}\left(d\right)$}{
			$s\leftarrow S_{e}\left[l\right]\left(\bfy[0],\bfy[1],\bfx[1],\bfx[0],d/c\right)$\;
			$\dot{s}\leftarrow \dot{ S_{e}}\left[l\right]\left(\bfy[0],\bfy[1],\bfx[1],\bfx[0],d/c\right)$\;
			$I\left[i,j,l\right]\leftarrow I\left[i,j,l\right] +wfs $\;
			$\dot{I}\left[i,j,l\right]\leftarrow \dot{I}\left[i,j,l\right] +w\dot{f} s+wf \dot{s}$\;
		}
	}
\end{algorithm}

\begin{algorithm}[htpb]
	\caption{Estimating the boundary integral}\label{alg:boundary}
	\SetAlgoLined
	\KwData{scene, pixel index $\left(i,j\right)$, max bounce $k$}
	\KwResult{Rendered scene derivative temporal histogram of the $\left(i,j\right)$-th pixel $\dot{I}\left[i,j,\cdot\right]$}
	Sample $\left(\bfx_B,\homega_B\right)\leftarrow \mathbb{P}_{\text{boundaryRay}}$ \;
	$\bfx_L\leftarrow \text{rayTrace}\left(\bfx_B,-\homega_B\right)$ \;
	$\bfx_S\leftarrow \text{rayTrace}\left(\bfx_B,\homega_B\right)$ \;
	\If{$\bfx_L$ and $\bfx_S$ are valid}{
		$f_B \leftarrow G\left(\bfx_L,\bfx_S\right) {\calV}_{\partial \hat\Omega}\left(\bfx_L,\bfx_S\right)J_B\left(\bfx_B,\homega_B\right)/\mathbb{P}_{\text{boundaryRay}}\left(\bfx_B,\homega_B\right)$\;
		$d_B \leftarrow \eta\norm{\bfx_L-\bfx_S}$ \;
		$\left(f_S , d_S, \bfx_0, \bfx_1\right)\leftarrow \mathrm{EstimateSensorSubpath}\left(\bfx_S\right)$\;
		$\left(f_L , d_L, \bfy_0, \bfy_1\right)\leftarrow \mathrm{EstimateSourceSubpath}\left(\bfx_L\right)$\;
		$d\leftarrow d_L+d_B+d_S$ \;
		\For{$l\in S_{e}.\mathrm{range}\left(d\right)$}{
			$s\leftarrow S_{e}\left[l\right]\left(\bfy_0,\bfy_1,\bfx_1,\bfx_0,d/c\right)$\;
			$\dot{I}\left[i,j,l\right]\leftarrow \dot{I}\left[i,j,l\right] +f_L f_B f_S s$\;
		}
	}
\end{algorithm}

\end{document}